\documentclass[onecolumn]{IEEEtran}  

\usepackage{algpseudocode}
\usepackage{amsthm}
\usepackage{mathtools}  
\usepackage{color}
\usepackage{graphicx}
\usepackage{array,multirow}
\usepackage[table]{xcolor}
\usepackage{hhline}
\usepackage{stmaryrd}
\usepackage{caption}
\usepackage{subcaption}

\usepackage{algorithm}

\usepackage{tikz}
\usetikzlibrary{positioning}
\usepackage{varwidth}
\usepackage{cite}

\ifCLASSINFOpdf
   \usepackage{epstopdf}
\else
\fi

\usepackage{cuted}
\usepackage{amsmath,graphicx,bm}
\usepackage{array,multirow}
\usepackage{stfloats}
\usepackage{amssymb}
\usepackage{color}
\usepackage{bbm}
\usepackage{mathtools}

\newcommand{\appropto}{\mathrel{\vcenter{
  \offinterlineskip\halign{\hfil$##$\cr
    \propto\cr\noalign{\kern2pt}\sim\cr\noalign{\kern-2pt}}}}}

\begin{document}
%

\title{A multi-sensor multi-Bernoulli filter}
%
%
%
\author{\IEEEauthorblockN{Augustin-Alexandru Saucan,
Mark Coates and
Michael Rabbat}
\thanks{The authors are with the Department of Electrical and Computer Engineering, McGill University, Montreal, QC, Canada (e-mail: augustin.saucan@mail.mcgill.ca). 
\newline 
\indent This work was supported by PWGSC contract W7707-145675$/$001$/$HAL funded by Defence R\&D Canada }}
\newtheorem{Proposition}{Proposition}
\newtheorem{theorem}{Theorem}[section]
\newtheorem{lemma}[theorem]{Lemma}
\newtheorem{proposition}[theorem]{Proposition}
\newtheorem{corollary}[theorem]{Corollary}
\newtheorem{definition}[theorem]{Definition}
\renewcommand{\qedsymbol}{$\blacksquare$}

\newcommand{\esper}{\mathbb{{E}}}
\newcommand{\norm}[1]{\left\lVert#1\right\rVert} 
\newcommand{\card}[1]{\left\lvert#1\right\rvert} 
\newcommand{\intset}[1]{\llbracket 1, #1 \rrbracket }
\newcommand{\dotprod}[2]{ \langle #1 \text{,}\, #2 \rangle  }

\newcommand{\gate}[1]{\tilde{#1}}
\newcommand{\Gate}[1]{\widetilde{#1}}

\newcommand{\vect}[1]{\mathbf{#1}}
\newcommand{\mat}[1]{\mathbf{#1}}
\newcommand{\set}[1]{{#1}}
\newcommand{\gsc}[1]{\tilde{#1}}
\newcommand{\gvect}[1]{\mathbf{\tilde{#1}}}
\newcommand{\gmat}[1]{\mathbf{\widetilde{#1}}}
\newcommand{\gset}[1]{\widetilde{#1}}
\newcommand{\vectg}[1]{\bm{#1}}
\newcommand{\matg}[1]{\bm{#1}}
\newcommand{\setg}[1]{{#1}}
\newcommand{\hsc}[1]{\hat{#1}}
\newcommand{\hvect}[1]{\mathbf{\hat{#1}}}
\newcommand{\hmat}[1]{\mathbf{\hat{#1}}}
\newcommand{\hset}[1]{\hat{#1}}
\newcommand{\hvectg}[1]{\bm{\hat{#1}}}
\newcommand{\hmatg}[1]{\bm{\hat{#1}}}
\newcommand{\hsetg}[1]{\hat{#1}}
\newcommand{\bvect}[1]{\mathbf{\bar{#1}}}
\newcommand{\bmat}[1]{\mathbf{\bar{#1}}}
\newcommand{\bset}[1]{\bar{#1}}
\newcommand{\bvectg}[1]{\bm{\bar{#1}}}
\newcommand{\bmatg}[1]{\bm{\bar{#1}}}
\newcommand{\bsetg}[1]{\bar{#1}}

\newcommand{\tsc}[1]{\tilde{#1}}
\newcommand{\wtsc}[1]{\widetilde{#1}}
\newcommand{\tvect}[1]{\mathbf{\tilde{#1}}}
\newcommand{\tvectg}[1]{\bm{\tilde{#1}}}
\newcommand{\tmat}[1]{\mathbf{\widetilde{#1}}}
\newcommand{\tmatg}[1]{\bm{\widetilde{#1}}}
\newcommand{\tset}[1]{\widetilde{#1}}

\newcommand\numberthis{\addtocounter{equation}{1}\tag{\theequation}}

\maketitle

\begin{abstract}

In this paper we derive a multi-sensor multi-Bernoulli (MS-MeMBer) filter for multi-target tracking. Measurements from multiple sensors are employed by the proposed filter to update a set of tracks modeled as a multi-Bernoulli random finite set. An exact implementation of the MS-MeMBer update procedure is computationally intractable. We propose an efficient approximate implementation  by using a greedy measurement partitioning mechanism. The proposed filter allows for Gaussian mixture or particle filter implementations. Numerical simulations conducted for both linear-Gaussian and non-linear models highlight the improved accuracy of the MS-MeMBer filter and its reduced computational load with respect to the multi-sensor cardinalized probability hypothesis density filter and the iterated-corrector cardinality-balanced multi-Bernoulli filter especially for low probabilities of detection.

\end{abstract}

\begin{IEEEkeywords}
Random finite sets, multi-sensor multi-Bernoulli filter, multi-sensor and multi-target tracking.
\end{IEEEkeywords}

%
\IEEEpeerreviewmaketitle


\section{Introduction}
\label{sec:intro}
\IEEEPARstart{S}{ingle} sensor multi-target tracking has received a great amount of attention in the scientific literature. Whenever the number of targets is unknown and time varying, a popular solution builds on the Random Finite Set (RFS) theory \cite{bibl:mahler_book2007}. In this category, the most well known filter is the Probability Hypothesis Density (PHD) filter of \cite{bibl:mahler_PHD2003}. The PHD filter models the multiple targets as a Poisson RFS, where the number of targets is Poisson distributed and the target distributions are independent and identically distributed (iid). The PHD filter adaptively estimates a function defined over the single-target space which is referred to as the PHD function. The number of targets and their states are inferred from the PHD function.

A different choice involves modeling each target as a Bernoulli RFS, characterized by a probability of existence and a target probability density. Accordingly, a set of independent targets is modeled by a multi-Bernoulli RFS, i.e., a union of independent Bernoulli RFSs. The multi-Bernoulli (MeMBer) filter was proposed in \cite[Ch. 17]{bibl:mahler_book2007} with subsequent improvements in \cite{bibl:vo_CMBF2009} and \cite{bibl:kiruba_member_IET2016}.

The multi-sensor scenario involves processing observations made by several sensors which are usually assumed to be conditionally independent given the target states. A generalized PHD filter for the special case of two sensors was first proposed in \cite{bibl:mahler_ms1_2009} and \cite{bibl:mahler_ms2_2009}, and extended in \cite{bibl:delade_ms_phd_report} and \cite{bibl:delande_ms_ICASP2011} to an arbitrary number of sensors. Approximate multi-sensor filters were developed in order to reduce the combinatorial complexity of the generalized PHD filter. Subsequently, in \cite{bibl:mahler_ms2_2009} the \textit{iterated-corrector} PHD filter was proposed and the \textit{approximate product multi-sensor} PHD and CPHD filters were introduced in \cite{bibl:mahler_msCPHD_fusion2010} and implemented in \cite{bibl:clark_ospa_fusion2011}. A comprehensive review of several of the aforementioned multi-sensor solutions is presented in \cite[Ch. 10]{bibl:mahler_adv_book_2014}. Other filters, such as \cite{bibl:papi_MdeltaGLMB_SPL2016} and \cite{bibl:extended_ms_dGLMB_2015}, rely on the $\delta$-Generalized Labeled multi-Bernoulli RFS in order to achieve approximate multi-sensor multi-target tracking. A generalized multi-sensor CPHD (MS-CPHD) filter was proposed in \cite{bibl:nannuru_MS_CPHD_2016} along with computationally-tractable implementations. In a different setting, in \cite{bibl:reza_fusion2016, bibl:reza_dist_member_GCI_TSP2017} distributed multi-target filtering is achieved using an unlabeled version of the Generalized Labeled multi-Bernoulli RFS and the Generalized Covariance Intersection method \cite{bibl:mahler_GCI_spie2000, bibl:clark_julier_GCI2010} for the fusion of posterior densities. 



The update process of the MS-PHD and MS-CPHD filters is achieved via Bayes' theorem and involves partitioning all of the sensor measurements into disjoint subsets. Each subset includes at most one measurement per sensor and corresponds to the measurements made by a potential target across all sensors. The subsets of a partition are disjoint and comprise all the sensor measurements. Exploring all partitions and subsets is impractical, and tractable implementations of the MS-PHD and MS-CPHD filters consider only the subsets (and subsequently the partitions) that make a significant contribution to the predicted PHD function. Hence, a greedy partitioning mechanism is employed which associates likely measurement subsets with individual target densities from the predicted PHD function. The MS-PHD and MS-CPHD filter implementations of \cite{bibl:nannuru_MS_CPHD_2016} constrain the PHD function to be a Gaussian mixture and assume that each Gaussian component represents a potential target.


In this paper we propose and derive a multi-Bernoulli filter for the multi-sensor case. The multi-Bernoulli RFS models each target as a separate Bernoulli RFS with its own probability density function and existence probability. Effectively, the multi-Bernoulli prior shifts the combinatorial problem to that of associating observation subsets to Bernoulli components. Therefore no clustering operations are required for measurement partitioning, and the probability density of each target (i.e., each Bernoulli component) can take whatever form is best suited for the target state model.
Furthermore, the proposed implementation of the MS-MeMBer filter has a simplified update procedure that reduces the computational complexity of the filter. More precisely, each Bernoulli component is only updated with its associated observation subsets; that is, subsets that have a significant contribution to the probability of existence of that component.

This paper is organized as follows. Section \ref{sec:back_rfs} reviews background information on RFS theory and introduces notation employed throughout the paper. Section \ref{sec:ss_member} presents an overview of single sensor multi-Bernoulli filtering. The proposed filter is derived in Section \ref{sec:ms_member} with numerical implementations being discussed in Section \ref{sec:implementation}. Then simulation results are presented for both linear-Gaussian (Section \ref{sec:lin_gauss_res}) and non-linear (Section \ref{sec:non_lin_res}) state systems. We conclude in Section \ref{sec:conclusions}.

\section{Random Finite Set Statistics}
\label{sec:back_rfs}
Throughout this paper, an RFS is employed to model a random number of targets with random state vectors. The realization of an RFS is a set $\set{X} = \lbrace \mathbf{x}_{1}, \dots, \mathbf{x}_{n} \rbrace$, where $n\geq 0$ is the random number of targets and $\mathbf{x}_{i}$ is the state vector of the $i$-th target. The cardinality of a finite set $\set{X}$, i.e., the number of elements is denoted with $\card{\set{X}}$. State vectors take values in the single-target space, $\mathbf{x} \in \mathbb{X}$, which is usually a subspace of $\mathbb{R}^d$. The random nature of an RFS is captured by its probability density $\pi(\set{X})$. The set of all finite subsets of $\mathbb{X}$ is denoted with $\mathcal{F}(\mathbb{X})$ and for a function $f: \mathcal{F}(\mathbb{X}) \rightarrow \mathbb{R}$, the set integral is defined as \cite[Ch. 11.3.3.1]{bibl:mahler_book2007}
\begin{equation}
\int f(\set{X}) \delta \set{X} \triangleq \sum_{n=0}^{\infty} \frac{1}{n!} \idotsint_{\mathbb{X}^n} f(\lbrace \vect{x}_1, \cdots, \vect{x}_n\rbrace)d\vect{x}_1 \cdots d\vect{x}_n.  
\label{eq:set_int}
\end{equation}
Additionally, we employ the exponential notation $u^{\set{X}} = \prod_{\vect{x} \in \set{X}} u(\vect{x})$, with $u^{\emptyset} = 1$ by convention. By introducing the test function $u:\mathbb{X} \rightarrow [0,1]$, the Probability Generating Functional (PGFl) \cite[Sec. 11.3.5]{bibl:mahler_book2007} is defined as 
\begin{equation}
G[u] = \int  u^{\set{X}} \pi(\set{X}) \delta \set{X}.
\label{eq:pgfl_def}
\end{equation}

The Bernoulli RFS is either an empty set with probability $1-r$ or a singleton set with probability $r$. In the latter case, the singleton is distributed according to a pdf $p(\vect{x})$, which represents the density of a single target. The PGFl of a Bernoulli RFS is given by (see \cite[pp. 374]{bibl:mahler_book2007})
\begin{equation}
G[u] = 1-r + r \dotprod{p}{u},
\label{eq:pgfl_ber_def}
\end{equation}  
\noindent where $\dotprod{p}{u} \triangleq \int p(\vect{x}) u(\vect{x}) d\vect{x} $ is the inner product. A multi-Bernoulli RFS is obtained by taking the union of $M$ independent Bernoulli RFSs, and its PGFl is
\begin{equation}
G[u] = \prod_{i=1}^{M} \left( 1 - r^{(i)} + r^{(i)} \dotprod{p^{(i)}}{u} \right),
\label{eq:pgfl_mber_def}
\end{equation}  
\noindent where $(r^{(i)}, p^{(i)})$ are the parameters of the $i$-th Bernoulli component. 

The functional derivative of a functional $F[u]$ in the direction of the Dirac delta density $\delta_{\vect{x}}$ is defined as $ \frac{\delta F}{\delta \vect{x}}[u] \triangleq  \frac{\partial F}{\partial \delta_{\vect{x}}}[u] = \text{lim}_{\epsilon \searrow 0} \frac{F[u + \epsilon \delta_{\vect{x}}] - F[u]}{\epsilon} $ (see \cite[Eq. 11.186]{bibl:mahler_book2007}). The first-order moment $D_{ }(\cdot)$ associated with $\pi_{ }(\cdot)$, also called the \emph{probability hypothesis density function}, is given by
\begin{align}
D_{}(\vect{x}) &= \left. \frac{\delta G_{ } }{\delta \vect{x}}[u]  \right\vert_{u(\vect{x})=1,\; \forall \, \vect{x}}
\label{eq:phd_def}
\\ &= \sum_{i=1}^{M_{} }r_{}^{(i)}\: p_{ }^{(i)}(\vect{x}).
\label{eq:multi_bernoulli_phd}
\end{align} 

\section{Single sensor multi-Bernoulli filters}
\label{sec:ss_member}
Several single sensor multi-Bernoulli filters have been proposed in the literature. In this section, we present an overview of several multi-Bernoulli filters with an emphasis on the Cardinality Balanced MeMBer (CBMeMBer) filter \cite{bibl:vo_CMBF2009}. The set of targets is modeled by a multi-Bernoulli RFS  $\set{X}_k$, indexed by the sample time $k$. The targets are observed by a single sensor that generates a set of measurements $\set{Z}_{k} = \lbrace \vect{z}_{k}^l \vert l =1, \dots, m_{k} \rbrace$ that contains at most one measurement per target and clutter measurements. The aim of all single sensor multi-Bernoulli filters is to provide an estimate of the posterior density $\pi_{k+1 \vert k+1}(\cdot)$ of the RFS $X_{k+1} $ given the set of all measurements $Z_{1:k+1} = \lbrace Z_1, \cdots, Z_{k+1}\rbrace$ up to and including time $k+1$.

We consider at time $k$ a multi-Bernoulli RFS with parameters $\lbrace (r_{k \vert k}^{(i)}, p_{k \vert k}^{(i)}) \rbrace_{i=1}^{M_{k \vert k}}$ and with posterior distribution $\pi_{k \vert k}(\cdot)$. The target kinematic model, birth, and death of targets are incorporated into the prediction stage of the filter. The correction of the predicted distribution, via the current sensor measurement set $\set{Z}_{k}$, is achieved in the update stage. We give the details of each stage next. 

\subsection{Single sensor MeMBer prediction}
\label{sec:ss_pred}
 
\sloppy Target death/disappearance is incorporated via the target probability of survival $p_{S,k}(\vect{x})$, and births are accounted for by appending a birth multi-Bernoulli RFS with components $ \lbrace ( r_{B, k+1}^{(i)},  p_{B, k+1 }^{(i)} ) \rbrace_{i=1}^{M_{B,k+1}}$ to the surviving targets. The birth Bernoulli RFSs are mutually independent and independent from the surviving targets. Additionally, the target kinematic model is incorporated via the transition kernel $f_{k+1 \vert k}(\vect{x}_{k+1} \vert \vect{x}_{k})$. The resulting RFS density $\pi_{k+1 \vert k}(\set{X})$ is multi-Bernoulli (see \cite{bibl:vo_CMBF2009}) and is comprised of the components  
\begin{equation}
\big\{ ( r_{k+1 \vert k}^{(i)},  p_{k+1 \vert k}^{(i)} ) \big\}_{i=1}^{M_{k+1 \vert k}} =  \big\{ ( r_{P,k+1 \vert k}^{(i)},  p_{P,k+1 \vert k}^{(i)} ) \big\}_{i=1}^{M_{k \vert k}}  \cup \big\{ ( r_{B, k+1}^{(i)},  p_{B, k+1 }^{(i)} ) \big\}_{i=1}^{M_{B,k+1}}, 
\label{eq:pred_components}
\end{equation} 
\noindent where the surviving Bernoulli components have parameters
 \begin{subequations}
\begin{align}
r_{P,k+1 \vert k}^{(i)} &= r_{k\vert k}^{(i)} \dotprod{p_{k \vert k}^{(i)}}{p_{S,k+1}}, 
\label{eq:up_func_dev1}
\\ p_{P,k+1 \vert k}^{(i)}(\vect{x})  &=  \frac{\dotprod{f_{k+1 \vert k}(\vect{x}\vert \cdot)}{ p_{k \vert k}^{(i)}p_{S,k+1}}}{ \dotprod{p_{k\vert k}^{(i)}}{p_{S,k+1}} }. 
\label{eq:up_func_dev2}
\end{align}
\end{subequations}  
\noindent Since the predicted RFS is a multi-Bernoulli RFS, it admits a PGFl of the form (\ref{eq:pgfl_mber_def}).


\subsection{Single sensor MeMBer update}
\label{sec:ss_up}

Let $\pi_{k+1 \vert k}(\set{X})$ be the density of the predicted RFS with PGFl given by (\ref{eq:pgfl_mber_def}). Denoting with $L_{k+1}(Z\vert X)$ the multi-target likelihood function of the measurement set $Z$ given the target set $X$, we define the functional $F[g,u]$ \cite[Eq. 14.281]{bibl:mahler_book2007} as
\begin{equation}
F[g,u] = \int u^X \left[ \int g^Z L_{k+1}(Z \vert X) \delta Z \right] \pi_{k+1 \vert k}(X) \delta X.
\label{eq:ss_pgfl_bi}
\end{equation}

The PGFl corresponding to the updated density $\pi_{k+1 \vert k+1}(\cdot)$ (i.e., corrected with the measurement set $Z_{k+1}$) is \cite[Sec. 14.8.2]{bibl:mahler_book2007}   
\begin{equation}
G_{k+1 \vert k+1}[u] = \frac{\frac{\delta F}{\delta Z_{k+1}}[0,u]}{\frac{\delta F}{\delta Z_{k+1}}[0,1]},
\label{eq:ss_pgfl_up}
\end{equation}
where $\frac{\delta F}{\delta Z_{k+1}}[g,u]$ is the functional derivative of $F$ in $g$ with respect to the set $Z_{k+1}$\cite[Eq. 11.191]{bibl:mahler_book2007}\footnote{For a general functional $H$, the functional derivative $\frac{\delta H}{\delta Y}[h] = H[h]$ for $Y =\emptyset$ and $\frac{\delta H}{\delta Y}[h]=\frac{\delta^n H}{\delta \vect{y}_1 \cdots \delta \vect{y}_n}$ when $Y = \lbrace \vect{y}_1 \cdots \vect{y}_n \rbrace$.}.    

As shown in \cite[Sec. 17.4.2]{bibl:mahler_book2007}, the PGFl $G_{k+1 \vert k+1}[u]$ does not have the form of (\ref{eq:pgfl_mber_def}), i.e., the updated posterior density $\pi_{k+1 \vert k+1}(\cdot)$ does not correspond to a multi-Bernoulli RFS. Therefore, in \cite[Sec. 17.4.2]{bibl:mahler_book2007} several approximations are applied to (\ref{eq:ss_pgfl_up}) in order to obtain a multi-Bernoulli approximation to $G_{k+1 \vert k+1}[u]$. The resulting filter is referred to as the MeMBer filter. However, in \cite{bibl:vo_CMBF2009} it is shown that the MeMBer filter has a positive cardinality bias, and an unbiased filter called the Cardinality Balanced MeMBer (CBMeMBer) filter is proposed. The CBMeMBer filter also employs several approximations. First, in the PGFl of (\ref{eq:ss_pgfl_up}) it is assumed that the clutter density is not too large. Second, a first-order moment (PHD approximation) is employed in order to obtain a multi-Bernoulli PGFl that captures the intensity function (and hence mean cardinality) of the original PGFl. Finally, the approximation of high probability of detection $p_{D,k} \approx 1$ is required in order to correct a negative term appearing in the probabilities of existence of the resulting multi-Bernoulli components. More details regarding the derivation of the CBMeMBer filter are given in \cite{bibl:vo_CMBF2009}. In \cite{bibl:kiruba_member_IET2016}, the cardinality bias of the MeMBer filter is alleviated by modeling spurious targets arising from the legacy track set. However, the resulting unbiased MeMBer filter also employs a low density clutter approximation.    

A straightforward extension of any of the single sensor MeMBer filters to the case of multiple sensors can be achieved by iterating the filter update stage for each sensor measurement set. For example, the filter obtained by sequentially processing the measurement set of each sensor with the CBMeMBer filter corrector leads to the Iterated-Corrector CBMeMBer (IC-CBMeMBer) filter.           

%
%
%
%

\section{Multi-sensor multi-Bernoulli (MS-MeMBer) Filter}
\label{sec:ms_member}

We consider a set of targets at time $k$ modeled as a multi-Bernoulli RFS and characterized by a posterior distribution with parameters $\lbrace (r_{k \vert k}^{(i)}, p_{k \vert k}^{(i)}) \rbrace_{i=1}^{M_{k \vert k}}$. The targets are observed by $s$ sensors which, conditional on the target states, generate independent measurements. Let $\set{Z}_{j,k} = \lbrace \vect{z}_{j,k}^1, \dots, \vect{z}_{j,k}^{m_{j,k}} \rbrace$ be the set of all measurements $\vect{z}_{j,k} \in \mathbb{Z}_{j}$ of the $j$-th sensor, with $\mathbb{Z}_j$ being the measurement space of sensor $j$. Let us also denote the collection of measurements collected by all sensors at time $k$ by $\set{Z}_{1:s,k} = (\set{Z}_{1,k}, \cdots, \set{Z}_{s,k})$. Again, we assume that each sensor can collect at most one measurement per target and that clutter measurements may be present. Multi-sensor multi-target filters provide an estimate of the distribution $\pi_{k+1 \vert k+1}(\cdot) $ of $\set{X}_{k+1}$ given $\set{Z}_{1:s, 1:k+1} $, which is obtained in a Bayesian framework via prediction and update. 


Similar to the single sensor case, the updated PGFl $G_{k+1 \vert k+1}[u]$, and subsequently the identification of updated multi-Bernoulli components, can be achieved via the differentiation of the multivariate functional $F[g_{1:s},u] \triangleq F[g_1,\dots, g_s,u]$, where the variable $g_i$ corresponds to the sensor $i$. We denote with $L_{i,k+1}(\set{Z}_{i} \vert \set{X})$ the multi-target likelihood function for sensor $i$ at time $k+1$. Considering the sensor measurements as conditionally independent given the multi-target state, and analogous to (\ref{eq:ss_pgfl_bi}), $F[g_{1:s},u]$ is defined as 
\begin{align}
  F[g_{1:s},u]  \triangleq 
  \int u^{\set{X}} \left( \prod_{i=1}^{s} \int g_i^{\set{Z}_{i}} L_{i,k+1}(\set{Z}_{i} \vert \set{X}) \delta \set{Z}_i \right) \pi_{k+1 \vert k}(\set{X} ) \delta \set{X}. 
\label{eq:up_functional_def}
\end{align} 
The parameters of the measurement model for sensor $i$ are the probability of detection $p_{i,D,k+1}(\cdot)$, likelihood function $h_{i,k+1}(\cdot)$, clutter probability density function (pdf) $c_{i,k+1}(\cdot)$, clutter cardinality distribution $p_{c,i,k+1}(n)$ and probability generating function (pgf) $ C_{i,k+1}(u)\triangleq \sum_{n=0}^{\infty} u^n p_{c,i,k+1}(n)$. 

From \cite[Sec. 12.3.7]{bibl:mahler_book2007}, the measurement PGFl can be written as
\begin{equation}
\int g_i^{\set{Z}_{i}} L_{i,k+1}(\set{Z}_{i} \vert \set{X}) \delta \set{Z}_i   = C_{i,k+1} (\left\langle c_{i,k+1}, g_i \right\rangle ) 
 \prod_{\vect{x}\in X} \bigg[ 1- p_{i,D,k+1}(\vect{x}) +  p_{i,D,k+1}(\vect{x})\int g_i(\vect{z}) h_{i,k+1}(\vect{z}\vert \vect{x})d\vect{z} \bigg]. 
\label{eq:obs_pgf}
\end{equation} 
\noindent In the following, for conciseness we omit the time index $k+1$ when it is clear from the context. 

We denote by $G^{(j)}_{k+1 \vert k} [\cdot]$ the PGFl of the $j$-th Bernoulli component of the predicted density $\pi_{k+1 \vert k}(\cdot)$. Note that $G^{(j)}_{k+1 \vert k} [\cdot]$ has the form of equation (\ref{eq:pgfl_ber_def}). Additionally we define the function
 \begin{equation}
\phi_{g_i}(\vect{x}) \triangleq 1 -p_{i,D}(\vect{x}) + p_{i,D}(\vect{x}) \int g_i(\vect{z})h_i(\vect{z} \vert \vect{x}) d\vect{z}. 
\label{eq:phi_func_def}
\end{equation}
\noindent Based on the specific form of (\ref{eq:obs_pgf}), the functional $F[g_{1:s},u]$ can be written as
\begin{align}
&F[g_{1:s},u]  \nonumber \\
&= \left( \prod_{i=1}^s C_i(\left\langle c_i,g_i \right\rangle ) \right) \int \left( u \textstyle{\prod_{i=1}^s} \phi_{g_i} \right)^X \pi_{k+1 \vert k}(X) \delta X, \nonumber \\
&= \left( \prod_{i=1}^s C_i(\left\langle c_i,g_i \right\rangle ) \right) \prod_{j=1}^{M_{k+1 \vert k}} G^{(j)}_{k+1 \vert k} [u \textstyle{\prod_{i=1}^s} \phi_{g_i}]. \numberthis \label{eq:up_func_dev2}
\end{align} 



\noindent Analogous to (\ref{eq:ss_pgfl_up}), the multi-sensor updated PGFl is given by
\begin{equation}
G_{k+1 \vert k+1}[u] = \frac{\frac{\delta^s F}{\delta \set{Z}_{1,k+1} \cdots \delta \set{Z}_{s,k+1}}[0,0,\dots,0,u]}{\frac{\delta^s F}{\delta \set{Z}_{1,k+1} \cdots \delta \set{Z}_{s,k+1}}[0,0,\dots,0,1]},
\label{eq:up_pgfl_def}
\end{equation} 
\noindent where the functional $F[g_{1:s},u]$ is differentiated in $g_1$ with respect to $Z_{1,k+1}$, in $g_2$ with respect to $Z_{2,k+1}$ and so on. 

The result of the differentiation in (\ref{eq:up_pgfl_def}) requires the partitioning of the measurements $\set{Z}_{1:s,k+1}$. Therefore, we introduce notation similar to \cite{bibl:nannuru_MS_CPHD_2016}. Let $\set{W}_i \subset \set{Z}_{i,k+1}$ be a measurement subset that contains at most one measurement from sensor $i$, i.e., $\card{\set{W}_i} \leq 1$. Additionally, we construct the ordered collection of measurement subsets as $\set{W}_{1:s} \triangleq \left( \set{W}_1, \dots, \set{W}_s \right) $, which contains at most one measurement from each sensor. We denote the special case when $W_i = \emptyset $ $\forall$ $i=1, \dots, s$ by $\emptyset_{1:s}$. We refer to $W_{1:s}$ as a \emph{multi-sensor measurement subset}. Each multi-sensor subset $\set{W}_{1:s}$ can also be specified via the set of indices $T_{\set{W}_{1:s}} = \lbrace (i,l) \vert \vect{z}_i^l \in \set{W_i}, \: \forall i=1,\dots, s \rbrace$ that specify the sensor index $i$ as well as the measurement index $l$. We say that two multi-sensor subsets $W_{1:s}^j= ( \set{W}_1^j, \dots, \set{W}_s^j )$ and $W_{1:s}^p=( \set{W}_1^p, \dots, \set{W}_s^p )$ are disjoint if $W_i^j \cap W_i^p = \emptyset$ $\forall \: i=1,\dots, s$. 

Given a set of disjoint multi-sensor subsets $\set{W}_{1:s}^1, \dots, \set{W}_{1:s}^n$, we define the collection of clutter measurements as $\set{W}^0_{1:s} = \left( W^0_1, \dots, W^0_s \right)$ with $ W^0_i = \set{Z}_{i,k+1} \setminus ( \cup_{j=1}^n \set{W}_i^j)$. Each $\set{W}_{1:s}^j$ for $j \neq 0$ can be interpreted as the collection of measurements of a specific target across all sensors and $\set{W}^0_{1:s}$ as the collection of clutter points. For a given $M$, we define a \textit{quasi-partition} $P$ of the measurements $Z_{1:s, k+1}$ as $P = (W^0_{1:s}, \cdots, W^M_{1:s},)$. Note that $\set{W}_{1:s}^1, \dots, \set{W}_{1:s}^M$ and $\set{W}_{1:s}^0$ are disjoint. We refer to $P$ as a \emph{measurement quasi-partition} since its elements $W_{1:s}$ are allowed to be empty, i.e., $W_{1:s} = \emptyset_{1:s}$. For a number $M$ of targets, the quasi-partition $P$ can be interpreted as a partitioning of the measurements $Z_{1:s, k+1}$ into target-originated multi-sensor subsets $W_{1:s}^j$ (one for each target) and the clutter subset $W^0_{1:s}$. Additionally, let $\mathcal{P}$ denote the set of all quasi-partitions $P$.

The result of the differentiation of (\ref{eq:up_func_dev2}) is summarized in the next lemma, and the proof is presented in Appendix \ref{app:proof_deriv_func}.
\begin{lemma} \label{the:derivated_funcl}
We define $\gamma(\vect{x}) \triangleq \prod_{i=1}^s \left[1-p_{i,D}(\vect{x}) \right]$ and for any multi-sensor subset $W^j_{1:s}$ we introduce the multi-sensor likelihood for a single target with state $\vect{x}$ as 
\begin{equation}
f(W^j_{1:s} \vert \vect{x}) \triangleq \;  \prod_{ \mathclap{(i,l) \in T_{W^j_{1:s}}}} \; \frac{p_{i,D}(\vect{x}) h_i(\vect{z}_i^{l} \vert \vect{x})}{c_i(\vect{z}_i^l)} \; \prod_{  \mathclap{(i,*) \notin T_{W^j_{1:s}} }}\; (1-p_{i,D}(\vect{x})).
\label{eq:my_ms_lik}
\end{equation}
\noindent For each $j=1,\dots, M_{k+1 \vert k}$, we define the functionals 
\begin{equation}
\varphi_{W_{1:s}^j}^j[ u] \triangleq  \begin{cases} 
1- r_{k+1 \vert k}^{(j)} + r_{k+1 \vert k}^{(j)} \dotprod{p_{k+1 \vert k}^{(j)}}{u \gamma}  , & \mbox{if } W^j_{1:s} = {\emptyset}_{1:s} 
\\  r_{k+1 \vert k}^{(j)} \int u(\vect{x}) p_{k+1 \vert k}^{(j)}(\vect{x}) f(W_{1:s}^j \vert \vect{x}) d\vect{x}, & \mbox{otherwise. } 
\end{cases}
\label{eq:varphi_def}
\end{equation}
Additionally, let $\Gamma_{i}\triangleq \prod_{ \vect{z} \in Z_{i,k+1}} c_i(\vect{z})$ and $ \mathcal{K}_{P} \triangleq \prod_{i=1}^s C_i^{(\card{W^0_i})} (0)$ with $C_i^{(n)}(\cdot)$ denoting the $n$-th derivative of the clutter probability generating function $C_i(\cdot)$. Then the differentiation of the functional (\ref{eq:up_func_dev2}) with respect to all the sensors evaluated at $(0,\dots,0)$ is
\begin{equation}
\frac{\delta^s F}{\delta Z_{1,k+1} \cdots \delta Z_{s,k+1}}[0,\dots,0,u] =   
\left[ \prod_{i=1}^s\Gamma_i \right]\sum_{ P \in \mathcal{P}} \mathcal{K}_{P} \left[ \prod_{j=1}^{M_{k+1 \vert k}} \varphi_{W_{1:s}^j}^j[u] \right]. 
 \label{eq:func_deriv_g_def}
\end{equation}
\end{lemma}
Lemma \ref{the:derivated_funcl} gives the numerator of (\ref{eq:up_pgfl_def}), while the denominator of (\ref{eq:up_pgfl_def}) is obtained by evaluating $\frac{\delta F}{\delta Z_{1:s,1:k}}[0,\dots,0,u]$ at $u(\vect{x}) = 1$. Because of the additional sum in (\ref{eq:func_deriv_g_def}), the PGFl of the updated posterior is not a multi-Bernoulli PGFl, i.e., a product of Bernoulli PGFls as in (\ref{eq:pgfl_mber_def}). Note that in order to achieve a multi-Bernoulli posterior, the derivation of the single sensor MeMBer filters involves approximating the derivative of $F[g,u]$ with respect to $g$ at the measurement set $Z_{1,k+1}$. In contrast, the result in (\ref{eq:func_deriv_g_def}) and $G_{k+1 \vert k+1}[u]$ are exact, and in the following we apply a single first-order multi-target moment approximation (similar to the PHD filter of \cite{bibl:mahler_PHD2003}) after all sensor measurements have been taken into account in the PGFl $G_{k+1 \vert k+1}[u]$.

We approximate the updated posterior with a multi-Bernoulli distribution of equal first-order moment (i.e. PHD function). We aim to construct a multi-Bernoulli RFS $\hat{\pi}_{k+1 \vert k+1}(\cdot)$ with identical PHD function to that of $\pi_{k+1 \vert k+1}(\cdot)$. Implicitly, $\hat{\pi}_{k+1 \vert k+1}(\cdot)$ and $\pi_{k+1 \vert k+1}(\cdot)$ have the same mean cardinality. The PHD function is summarized in the following theorem, and its proof is presented in Appendix \ref{app:proof_updated_phd}.

\begin{theorem} \label{the:ms_mber_phd}
By defining the coefficients 
\begin{equation}
\alpha_{P} \triangleq \frac{ \mathcal{K}_{P} \displaystyle  \prod_{j=1}^{M_{k+1 \vert k}} \varphi_{W_{1:s}^j}^{j}[1] }{ \displaystyle \sum_{{Q} \in \mathcal{P}}  \mathcal{K}_{{Q}} \prod_{ j=1  }^{M_{k+1 \vert k}}   \varphi_{W_{1:s}^j}^{j}[1]} 
\label{eq:alpha_def}
\end{equation}
and the function 
\begin{equation}
\rho_{W_{1:s}^j}^j(\vect{x})  =  \begin{cases} 
\frac{r_{k+1 \vert k}^{(j)} \gamma(\vect{x})}{ 1- r_{k+1 \vert k}^{(j)} + r_{k+1 \vert k}^{(j)} \dotprod{p_{k+1 \vert k}^{(j)}}{\gamma}}  , & \mbox{if } W^j_{1:s} = \emptyset_{1:s} 
\\  \frac{f(W_{1:s}^j \vert \vect{x})}{ \int  p_{k+1 \vert k}^{(j)}(\vect{x}) f(W_{1:s}^j \vert \vect{x}) d\vect{x}}, & \mbox{otherwise,}
\end{cases}
\label{eq:rho_def_main}
\end{equation}
the PHD function obtained via (\ref{eq:phd_def}) from the PGFl $G_{k+1 \vert k+1}[u]$ is
\begin{equation}
D_{k+1 \vert k+1}(\vect{x}) =  \hspace{0.5cm}\mathclap{\sum_{\substack{P \in \mathcal{P} \\ P = \left( W_{1:s}^0, \dots, W_{1:s}^M \right)}}} \hspace{0.7cm}  \alpha_{P} \sum_{ j =1 }^{M_{k+1 \vert k}}   \rho_{W_{1:s}^j}^j(\vect{x}) p_{k+1 \vert k}^{(j)} (\vect{x}).
\label{eq:updated_phd}
\end{equation}
\end{theorem} 
The inner summation of (\ref{eq:updated_phd}) comprises the $M_{k+1 \vert k}$ predicted Bernoulli terms and effectively involves the update of each of the predicted Bernoulli components. For each quasi-partition $P = \left( W_{1:s}^0, \dots, W_{1:s}^M \right)$, the update process assigns the multi-sensor subset $\set{W}^i_{1:s}$ to one of the $M_{k+1 \vert k}$ predicted Bernoulli components. 
 
Theorem \ref{the:ms_mber_phd} shows that even though the PGFl $G_{k+1 \vert k+1}[u]$ is not multi-Bernoulli, its PHD function has a similar structure (i.e., a sum of weighted densities) as that of a multi-Bernoulli PHD [see eq. (\ref{eq:multi_bernoulli_phd})]. Therefore, the Bernoulli components of $\hat{\pi}_{k+1 \vert k+1}(\cdot)$ are identifiable from the expression in (\ref{eq:updated_phd}). A proposition for the Bernoulli components of $\hat{\pi}_{k+1 \vert k+1}(\cdot)$ is: 
\begin{equation}
\bigg\{ \left( \hat{r}_{k+1 \vert k+1}^{(j)}, \, \hat{p}_{k+1 \vert k+1}^{(j)}(\cdot) \right) \bigg\}_{j=1}^{\hat{M}_{k+1 \vert k+1}}   
= \displaystyle \bigcup_{ P \in \mathcal{P} }   \bigcup_{ j=1  }^{M_{k+1 \vert k}} \bigg\{  \left( r_{P}^{(j)}, \, p_{P}^{(j)}(\cdot) \right) \bigg\}, 
\label{eq:up_components}
\end{equation}
\noindent where
\begin{equation}
r_{P}^{(j)} = \begin{cases} 
{\alpha_{P}}  \frac{r_{k+1 \vert k}^{(j)}\dotprod{p_{k+1 \vert k}^{(j)}}{ \gamma} }{1- r_{k+1 \vert k}^{(j)} + r_{k+1 \vert k}^{(j)}\dotprod{p_{k+1 \vert k}^{(j)}}{ \gamma}}, & \mbox{if } W^j_{1:s} = \emptyset_{1:s}
\\  \alpha_{P}     & \mbox{otherwise }  
\end{cases}
\label{eq:up_components_def1}
\end{equation}
\noindent and
\begin{equation}
p_{P}^{(j)}(\vect{x}) = \begin{cases} 
\frac{p_{k+1 \vert k}^{(j)}(\vect{x}) \gamma(\vect{x})}{\dotprod{p_{k+1 \vert k}^{(j)}}{ \gamma}}, & \mbox{if } W^j_{1:s} = \emptyset_{1:s}.
\\  \frac{p_{k+1 \vert k}^{(j)}(\vect{x})f(W_{1:s}^j \vert \vect{x})}{ \int  p_{k+1 \vert k}^{(j)}(\vect{x}) f(W_{1:s}^j \vert \vect{x}) d\vect{x}}    , & \mbox{otherwise}. 
\end{cases}
\label{eq:up_components_def2}
\end{equation}

Note that the PHD function (\ref{eq:updated_phd}) is a mixture of densities, and the choice of updated Bernoulli RFSs is not unique. The choice made to arrive at (\ref{eq:up_components}) involves creating a Bernoulli RFS for each association between a predicted Bernoulli RFS and a multi-sensor subset. Other choices could be obtained by clustering several updated densities into a single Bernoulli component. A different proposal involves matching the cardinality distribution in addition to the intensity function of the approximate multi-Bernoulli RFS to the exact updated RFS. Alternatively, an approximating multi-Bernoulli density could be obtained by minimizing the Kullback-Leibler divergence from the exact RFS density. In \cite{bibl:williams_MeMBerVar_TSP2015}, this problem is solved via the expectation-maximization algorithm where the correspondence between the Bernoulli components in the best-fitting distribution and the components of the exact distribution are treated as missing data. However, this additional minimization step increases the computational complexity of the resulting algorithm and the development of efficient algorithms for the multi-sensor case are left for future investigation. 


\section{MS-MeMBer Practical implementation}
\label{sec:implementation}
The update process involves associating various measurement subsets $W_{1:s}$ with the predicted Bernoulli components without imposing any restrictions on the shape of the probability density of the Bernoulli components. Indeed, both Gaussian mixture 
 \begin{subequations}
\begin{align}
p_{k+1 \vert k}^{(j)}(\vect{x}) &= \sum_{n=1}^{J_{k+1}^{(j)}} w_{n,k+1}^{(j)} \; \mathcal{N}(\vect{x}; \vectg{\mu}_{n,k+1}^{(j)}, \matg{\Sigma}_{n,k+1}^{(j)}) 
\label{eq:gm_pdf}
\intertext{and particle based}
 p_{k+1 \vert k}^{(j)}(\vect{x}) &= \sum_{n=1}^{J_{k+1}^{(j)}} w_{n,k+1}^{(j)} \; \delta_{\vect{x}_{n,k+1}^{(j)}}(\vect{x})
\label{eq:pf_pdf}
\end{align}
\end{subequations}  
\noindent representations are possible. 
   
The challenge posed by the multi-Bernoulli density $\hat{\pi}_{k+1 \vert k+1}(\cdot)$ is given by the large number of Bernoulli components resulting after the update step (\ref{eq:up_components}). More precisely, starting with a number $M_{k+1 \vert k}$ of predicted Bernoulli components, defined in (\ref{eq:pred_components}), the total number of updated Bernoulli components is $ \sum_{P \in \mathcal{P}} M_{k+1 \vert k}$, the majority of which contribute very little to the updated PHD function. Therefore, a greedy mechanism for selecting the best associations and subsequently the best quasi-partitions is necessary. In the following, we show that $\varphi_{W^j_{1:s}}^j[1]$ (defined in (\ref{eq:varphi_def})) effectively scores the association of the measurement collection $W_{1:s}^j$ with the $j$-th Bernoulli component while $\alpha_{P}$ (defined in (\ref{eq:alpha_def})) represents a score of the quasi-partition $P$. Thus, similar to \cite{bibl:nannuru_MS_CPHD_2016} the scores $\varphi_{W^j_{1:s}}^j[1]$ and $\alpha_{P}$ can be employed to select high-scoring measurement subsets $W_{1:s}^j$, followed by high-scoring quasi-partitions $P$. The formation of measurement subsets $W_{1:s}$ and the formation of quasi-partitions are described in the following two sections. 
\subsection{Formation of multi-sensor subsets $W_{1:s}$}
\label{sec:subset_form}
\begin{figure}[!b]
\begin{center}
 {
\begin{tikzpicture}
\node [] (00) {\begin{varwidth}{1cm}\centering 1 \end{varwidth}};
\node [below=0.3cm of 00,draw,circle,minimum size=2.em,inner sep=1,line width=0.5mm] (01) {\begin{varwidth}{1cm}\centering $\emptyset$ \end{varwidth}};
\node [below=0.4cm of 01,draw,circle,minimum size=2.em,inner sep=1,line width=0.5mm] (02) {\begin{varwidth}{1cm}\centering $\vect{z}_1^1$ \end{varwidth}};
\node [below=-0.15cm of 02] (03) {\centering \textbf{$\vdots$}};
\node [below=-0cm of 03,draw,circle,minimum size=2.em,inner sep=1,line width=0.5mm] (04) {\begin{varwidth}{1cm}\centering $\vect{z}_1^{m_1}$ \end{varwidth}};
\node [right=1.2cm of 00] (10) {\begin{varwidth}{1cm}\centering 2 \end{varwidth}};
\node [below=0.3cm of 10,draw,circle,minimum size=2.em,inner sep=1,line width=0.5mm] (11) {\begin{varwidth}{1cm}\centering $\emptyset$ \end{varwidth}};
\node [below=0.4cm of 11,draw,circle,minimum size=2.em,inner sep=1,line width=0.5mm] (12) {\begin{varwidth}{1cm}\centering $\vect{z}_2^1$ \end{varwidth}};
\node [below=-0.15cm of 12] (13) {\centering \textbf{$\vdots$}};
\node [below=-0cm of 13,draw,circle,minimum size=2.em,inner sep=1,line width=0.5mm] (14) {\begin{varwidth}{1cm}\centering $\vect{z}_2^{m_2}$ \end{varwidth}};
\node [right=1.2cm of 10] (20) {\begin{varwidth}{1cm}\centering 3 \end{varwidth}};
\node [below=0.3cm of 20,draw,circle,minimum size=2.em,inner sep=1,line width=0.5mm] (21) {\begin{varwidth}{1cm}\centering $\emptyset$ \end{varwidth}};
\node [below=0.4cm of 21,draw,circle,minimum size=2.em,inner sep=1,line width=0.5mm] (22) {\begin{varwidth}{1cm}\centering $\vect{z}_3^1$ \end{varwidth}};
\node [below=-0.15cm of 22] (23) {\centering \textbf{$\vdots$}};
\node [below=-0cm of 23,draw,circle,minimum size=2.em,inner sep=1,line width=0.5mm] (24) {\begin{varwidth}{1cm}\centering $\vect{z}_3^{m_3}$ \end{varwidth}};
\node[right=0.3cm of 21] (x) {\centering $\boldsymbol{\cdots}$};
\node[right=0.3cm of 22] (x) {\centering $\boldsymbol{\cdots}$};
\node[right=0.3cm of 24] (x) {\centering $\boldsymbol{\cdots}$};
\node [right=1.7cm of 20] (30) {\begin{varwidth}{1cm}\centering $s$ \end{varwidth}};
\node [below=0.3cm of 30,draw,circle,minimum size=2.em,inner sep=1,line width=0.5mm] (31) {\begin{varwidth}{1cm}\centering $\emptyset$ \end{varwidth}};
\node [below=0.4cm of 31,draw,circle,minimum size=2.em,inner sep=1,line width=0.5mm] (32) {\begin{varwidth}{1cm}\centering $\vect{z}_s^1$ \end{varwidth}};
\node [below=-0.15cm of 32] (33) {\centering \textbf{$\vdots$}};
\node [below=0cm of 33,draw,circle,minimum size=2.em,inner sep=1,line width=0.5mm] (34) {\begin{varwidth}{1cm}\centering $\vect{z}_s^{m_s}$ \end{varwidth}};
\draw[->,color=green,line width=0.5mm,dashed] (12) -- (21);
\draw[->,color=blue,line width=0.5mm,dashed] (12) -- (22);
\draw[->,color=red,line width=0.5mm,dashed] (12) -- (24);
\node [above left=0.1cm of 20] (t1) {\begin{varwidth}{1cm}\centering  Sensors: \end{varwidth}};
\node [left=0.5cm of 02] (t2) {\begin{varwidth}{1cm}\centering  $W_{1:2}^{j}$ \end{varwidth}};
\draw[-,line width=0.5mm] (t2)-- (02);
\draw[-,line width=0.5mm] (02)-- (12);
\end{tikzpicture}}
\caption{Trellis diagram formed from the measurements of the $s$ sensors. For the $j$-th predicted Bernoulli component, a partial collection of measurements $W_{1:3}^j$ is formed by greedily appending high-scoring measurements from sensor $3$ to the previous partial collection $W_{1:2}^j$. The candidate measurements from sensor $3$ are evaluated through the score $\beta_{1:3}^j(W_{1:3}^j)$.}
\label{fig:subset_form}
\end{center}
\end{figure}
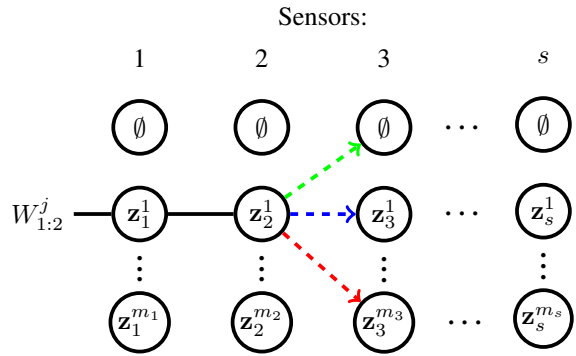
We employ $\varphi_{W^j_{1:s}}^j[1]$ as a measure of the likelihood that $W_{1:s}^j$ was generated by the $j$-th Bernoulli component. As we can see from (\ref{eq:my_ms_lik}) and (\ref{eq:varphi_def}), for non-empty multi-sensor subsets, $\varphi_{W^j_{1:s}}^j[1]$ can be interpreted as a ratio of the likelihood that $W_{1:s}^j$ was generated by the $j$-th Bernoulli component to the likelihood that $W_{1:s}^j$ is clutter. For the particular case of $W_{1:s}^j = \emptyset_{1:s} $, $\varphi_{W^j_{1:s}}^j[1]$ quantifies the probability that either all $s$ sensors have failed to detect the $j$-th Bernoulli component or the component no longer exists. We propose a greedy algorithm for selecting high-scoring measurement subsets $W_{1:s}$ for each predicted Bernoulli component by sequentially processing each sensor. The formation of $W_{1:p}^j$ for $p=1,\dots, s$ is depicted in Fig. \ref{fig:subset_form} as the formation of paths through the trellis formed with all sensor measurements and the empty measurement set (corresponding to the missed-detection case). A new measurement $\vect{z}_{p+1}^n$ is appended to an existing path $W_{1:p}^j$ if it maximizes the score $\beta_{1:p+1}^j(W_{1:p+1}^j) = \varphi^j_{W_{1:p+1}^j}[1]$. For the two representations of the pdf $p_{k+1 \vert k}^{(j)}(\cdot)$ in (\ref{eq:gm_pdf}) and (\ref{eq:pf_pdf}), the score $\beta_{1:p+1}^j(W_{1:p+1}^j)$ takes one of the forms
\begin{subequations}
\begin{align}
& r_{k+1 \vert k}^{(j)}\sum_{n=1}^{J_{k+1}^{(j)}} w_{n, k+1}^{(j)} \;\; \mathclap{ \int } \;\; \mathcal{N}(\vect{x}; \vectg{\mu}_{n,k+1}^{(j)}, \matg{\Sigma}_{n,k+1}^{(j)})  f(W_{1:p+1}^j \vert \vect{x})d\vect{x}
\label{eq:gm_score}
 \intertext{or} & r_{k+1 \vert k}^{(j)}  \sum_{n=1}^{J_{k+1}^{(j)}} w_{n,k+1}^{(j)} f(W_{1:{p+1}}^j\vert \vect{x}_{n,k+1}^{(j)}). 
\label{eq:pf_score}
\end{align}
\end{subequations}
In the Gaussian mixture case, the sensor detection probabilities are constant and the observation model is linear Gaussian, i.e. $h_i(\vect{z} \vert \vect{x}) = \mathcal{N}(\vect{z}; \mat{H}_i\vect{x}, \mat{R}_i)$ for some matrices $\mat{H}_i$ and $\mat{R}_i$ of appropriate dimensions. In the aforementioned conditions, (\ref{eq:gm_score}) admits an analytic form due to the properties of quadratic combinations \cite[App. 3.8]{bibl:beyondKalman}. 

In order to obtain the $n_s^j \geq 1$ best-scoring multi-sensor subsets for the $j$-th Bernoulli component, at each sensor we keep at most the highest $W_{\text{max}}$ scoring subsets. More precisely, starting with the set $\lbrace W_{1:p}^{j,i} \rbrace_{i=1}^{n_p^j}$ of subsets up to and including sensor $p$, we evaluate all possible extensions $(W_{1:p}^{j,i}, \{\vect{z}_{p+1}^l  \})$ with $\vect{z}_{p+1}^l \in Z_{p+1}$. A number $n_{p+1}^j \leq W_{\text{max}}$ of subsets with highest $\beta_{1:p+1}^j(W_{1:p+1}^{j,i})$ are selected in addition to the empty multi-sensor subset $\emptyset_{1:p+1} = (\emptyset, \dots, \emptyset)$. In this manner and after processing all $s$ sensors, we obtain $n_s^j\leq W_{\text{max}}$ high-scoring multi-sensor subsets in addition to $\emptyset_{1:s}$. Note that the maximum number of non-empty subsets $W_{\text{max}}$ is treated as a user-defined parameter in this work. A pseudo-code description of the greedy subset selection algorithm is given in Appendix \ref{app:subset_sel}. The computational complexity of the algorithm is $\mathcal{O}(M_{k+1\vert k}W_{\text{max}}\,\sum_{i=1}^s m_i)$, signifying a linear complexity with respect to the number of predicted Bernoulli components $M_{k+1 \vert k}$ and with the number of sensors $s$. Otherwise, note that an exhaustive enumeration of all $\prod_{i=1}^s (1+m_i)$ multi-sensor subsets and their associated scores for all predicted Bernoulli components involves a computational complexity of $\mathcal{O}(M_{k+1 \vert k}\prod_{i=1}^s m_i)$, which is exponential in the number of sensors $s$. In the following section, the greedily selected subsets are employed by the partition selection algorithm to form a number of high-scoring partitions.


\subsection{Formation of quasi-partitions $P$}
\label{sec:part_form}
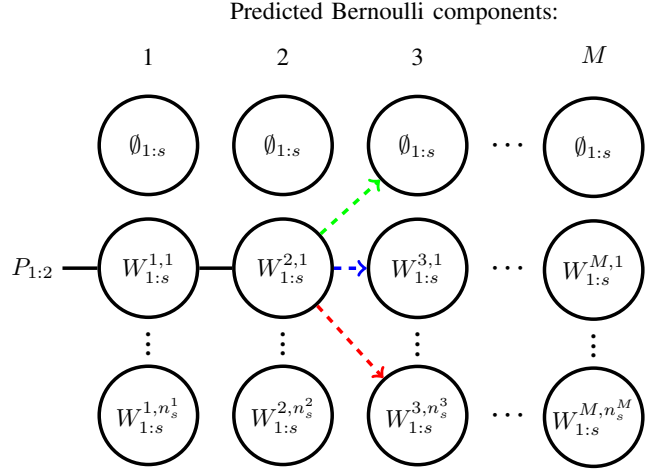
\begin{figure}[!t]

\begin{center}

 {\resizebox{0.47\textwidth}{!}{%

\begin{tikzpicture}

\node [] (00) {\begin{varwidth}{1cm}\centering 1 \end{varwidth}};

\node [below=0.3cm of 00,draw,circle,minimum size=4em,inner sep=1,line width=0.5mm] (01) {\begin{varwidth}{1.1cm}\centering $\emptyset_{1:s}$ \end{varwidth}};

\node [below=0.3cm of 01,draw,circle,minimum size=4em,inner sep=1,line width=0.5mm] (02) {\begin{varwidth}{1.1cm}\centering $W_{1:s}^{1,1}$ \end{varwidth}};

\node [below=-0.15cm of 02] (03) {\centering \textbf{$\vdots$}};

\node [below=0cm of 03,draw,circle,minimum size=4em,inner sep=1,line width=0.5mm] (04) {\begin{varwidth}{1.1cm}\centering $W_{1:s}^{1,n_s^1}$ \end{varwidth}};

\node [right=1.5cm of 00] (10) {\begin{varwidth}{1cm}\centering 2 \end{varwidth}};

\node [below=0.3cm of 10,draw,circle,minimum size=4em,inner sep=1,line width=0.5mm] (11) {\begin{varwidth}{1.25cm}\centering $\emptyset_{1:s}$ \end{varwidth}};

\node [below=0.3cm of 11,draw,circle,minimum size=4em,inner sep=1,line width=0.5mm] (12) {\begin{varwidth}{1.25cm}\centering $W_{1:s}^{2,1}$ \end{varwidth}};

\node [below=-0.15cm of 12] (13) {\centering \textbf{$\vdots$}};

\node [below=0cm of 13,draw,circle,minimum size=4em,inner sep=1,line width=0.5mm] (14) {\begin{varwidth}{1.25cm}\centering $W_{1:s}^{2,n_s^2}$ \end{varwidth}};

\node [right=1.5cm of 10] (20) {\begin{varwidth}{1cm}\centering 3 \end{varwidth}};

\node [below=0.3cm of 20,draw,circle,minimum size=4em,inner sep=1,line width=0.5mm] (21) {\begin{varwidth}{1.25cm}\centering $\emptyset_{1:s}$ \end{varwidth}};

\node [below=0.3cm of 21,draw,circle,minimum size=4em,inner sep=1,line width=0.5mm] (22) {\begin{varwidth}{1.25cm}\centering $W_{1:s}^{3,1}$ \end{varwidth}};

\node [below=-0.15cm of 22] (23) {\centering \textbf{$\vdots$}};

\node [below=0cm of 23,draw,circle,minimum size=4em,inner sep=1,line width=0.5mm] (24) {\begin{varwidth}{1.25cm}\centering $W_{1:s}^{3,n_s^3}$ \end{varwidth}};

\node[right=0.2cm of 21] (x) {\centering $\boldsymbol{\cdots}$};

\node[right=0.2cm of 22] (x) {\centering $\boldsymbol{\cdots}$};

\node[right=0.2cm of 24] (x) {\centering $\boldsymbol{\cdots}$};

\node [right=2cm of 20] (30) {\begin{varwidth}{1cm}\centering $M_{}$ \end{varwidth}};

\node [below=0.3cm of 30,draw,circle,minimum size=4em,inner sep=1,line width=0.5mm] (31) {\begin{varwidth}{1.25cm}\centering $\emptyset_{1:s}$ \end{varwidth}};

\node [below=0.3cm of 31,draw,circle,minimum size=4em,inner sep=1,line width=0.5mm] (32) {\begin{varwidth}{1.25cm}\centering $W_{1:s}^{M,1}$ \end{varwidth}};

\node [below=-0.15cm of 32] (33) {\centering \textbf{$\vdots$}};

\node [below=0cm of 33,draw,circle,minimum size=4em,inner sep=1,line width=0.5mm] (34) {\begin{varwidth}{1.25cm}\centering $W_{1:s}^{M,n_s^M}$ \end{varwidth}};

\draw[->,color=green,line width=0.5mm,dashed] (12) -- (21);

\draw[->,color=blue,line width=0.5mm,dashed] (12) -- (22);

\draw[->,color=red,line width=0.5mm,dashed] (12) -- (24);

\node [above left=0.1cm of 30] (t1) {\begin{varwidth}{7cm}\centering  Predicted Bernoulli components: \end{varwidth}};

\node [left=0.5cm of 02] (t2) {\begin{varwidth}{1cm}\centering  $P_{1:2}$ \end{varwidth}};

\draw[-,line width=0.5mm] (t2)-- (02);

\draw[-,line width=0.5mm] (02)-- (12);

\end{tikzpicture}}}

\caption{Trellis diagram formed with the multi-sensor subsets $W_{1:s}^j$ from the $j=1, \dots, M$ predicted Bernoulli components. Quasi-partitions are formed by sequentially processing the Bernoulli components and greedily appending the highest-scoring valid multi-sensor subsets.}

\label{fig:part_form}

\end{center}

\end{figure}

Quasi-partitions are formed as paths through a trellis constructed from the $n_s^{j}$ measurement subsets of the $j=1,\dots, M_{k+1\vert k}$ predicted Bernoulli components, as seen in Fig. \ref{fig:part_form}. Paths are formed sequentially across the $M_{k+1\vert k}$ Bernoulli components in a similar manner to the formation of multi-sensor subsets described in Section \ref{sec:subset_form}. Note from (\ref{eq:alpha_def}) that $\alpha_P$ serves to score a quasi-partition $P$. Hence, we define the score of a partial path $P_{1:p} \triangleq (W_{1:s}^1, \dots, W_{1:s}^p)$ as $\alpha_{1:p}(P_{1:p}) = \prod_{j=1}^p \beta^j_{1:s}(W_{1:s}^{j})$. Based on the scores $\alpha_{1:p+1}(\cdot) = \alpha_{1:p}(\cdot)  \beta^{p+1}_{1:s}(W_{1:s}^{p+1})$, a specific collection $W_{1:s}^{p+1}$ from the $p+1$ Bernoulli component is appended to the path $P_{1:p}$ in order to form $P_{1:p+1}=(W_{1:s}^1,\dots, W_{1:s}^{p+1})$. Thus, a complete path $P_{1:M_{k+1 \vert k}}$ is selected with the goal of maximizing $\alpha_{1:M_{k+1 \vert k}}(P_{1:M_{k+1 \vert k}})$, which is proportional to (\ref{eq:alpha_def}). 

Note that an additional operation needs to be performed in order to ensure that all multi-sensor subsets of a quasi-partition are pairwise disjoint. In other words, we need to ensure that the same measurement does not appear in two different multi-sensor subsets of the same quasi-partition. As noted in \cite[Sec. V.C]{bibl:nannuru_MS_CPHD_2016}, in the formation of a quasi-partition it is necessary to pre-select only valid multi-sensor subsets, i.e., those which are pairwise disjoint with the multi-sensor subsets in the current path. More specifically, considering the partial path $P_{1:p}$, we select the highest-scoring valid multi-sensor subsets from $\lbrace \emptyset_{1:s}, W_{1:s}^{p+1,1}, \dots, W_{1:s}^{p+1,n_s^{p+1}}\rbrace$. Note that the empty multi-sensor subset $\emptyset_{1:s}$ ensures that there will always exist a valid path through the trellis of Fig. \ref{fig:part_form}. In Algorithm~\ref{alg:part_sel} of Appendix~\ref{app:part_sel}, we present the pseudo-code for the greedy quasi-partition formation algorithm. The computational complexity of Algorithm \ref{alg:part_sel} is of $\mathcal{O}(P_{\text{max}}\,W_{\text{max}}\,s\, M_{k+1 \vert k}^2)$, involving a linear complexity with respect to the number of sensors but quadratic with respect to the number of predicted Bernoulli components. 


A maximal number of $P_{\text{max}}$ quasi-partitions, with the largest scores, are formed using a procedure similar to that employed to identify the $W_{\text{max}}$ subsets during the multi-sensor subset selection procedure. The updated Multi-Bernoulli RFS will contain at most $P_{\text{max}} M_{k+1\vert k}$ distinct Bernoulli components. In practice, different quasi-partitions might contain the same subset-to-Bernoulli assignment and thus create updated Bernoulli components with identical pdfs in (\ref{eq:up_components}) but with different probabilities of existence. Such components can be collapsed into a single Bernoulli component by adding together their existence probabilities. Thus, we obtain a density $\hat{\pi}_{k+1 \vert k+1}(\cdot)$ with a number of components $\hat{M}_{k+1 \vert k+1} \leq P_{\text{max}}M_{k+1\vert k}$ that has a PHD function approximately equal to (\ref{eq:updated_phd}).

In contrast with the partition selection procedure of the MS-CPHD filter \cite{bibl:nannuru_MS_CPHD_2016}, here, quasi-partitions are defined as ordered, i.e., $P = (W_{1:s}^0, \dots, W_{1:s}^{M_{k+1 \vert k}})$ and this implicitly ensures the association of the multi-sensor subset $W_{1:s}^j$ with the $j$-th Bernoulli component. In \cite{bibl:nannuru_MS_CPHD_2016}, each multi-sensor subset $W_{1:s}$ of a partition is employed to update all mixture components of the predicted PHD function. Following an approach similar to the quasi-partition selection mechanism presented above, a truncated MS-CPHD (MS-TCPHD) filter algorithm can be developed. In the case of the MS-TCPHD filter, a subset $W_{}$ from a partition is used to update only the $j$-th component of the PHD function, i.e., the component that maximizes the association score $\beta^{(j)}(W_{})$ (see equation (35) of \cite{bibl:nannuru_MS_CPHD_2016}). In in Appendix \ref{app:tcphd}, we present a detailed description of the MS-TCPHD filter. This truncated update mechanism of the MS-TCPHD filter is in contrast with the MS-CPHD filter, where each predicted PHD component is updated using all subsets of a partition.

\section{Numerical simulations}
\label{sec:results}
In this section we evaluate the performance of the proposed Multi-Sensor MeMBer (MS-MeMBer) filter with respect to the Multi-Sensor CPHD (MS-CPHD), the Multi-Sensor Truncated CPHD (MS-TCPHD) and the Iterated-Corrector Cardinality Balanced MeMBer (IC-CBMeMBer) filters. For simplicity the kinematic target model is linear and Gaussian. More specifically, we employ a white noise acceleration model described in \cite[Ch. 6.2.2]{bibl:BarShalomBook}. Regarding the measurement model, we consider two scenarios. The first involves a linear and Gaussian measurement equation so that the target state system becomes linear and Gaussian. In this case, we use Gaussian mixture implementations for all filters. The second scenario supposes a non-linear measurement equation (specifically, Doppler-bearing measurements) and consequently we rely on the unscented transform and SMC methods to implement the different filters. The scenarios aim to compare the performance of the filters and their computational times. Regarding the performance of filters, we employ the Optimal Sub-Pattern Assignment (OSPA) distance \cite{bibl:schuhmacher_ospaTSP2008} as the error metric. The OSPA metric measures both errors in the estimated number of targets as well as errors in state estimates of individual targets. The two parameters employed by OSPA are the cardinality penalty factor $c=100$ and order $p=1$. The simulations were performed using MATLAB \footnote{The code is available online at http://networks.ece.mcgill.ca/Augustin-Alexandru.Saucan.}.

\subsection{Target kinematic model}
\label{sec:target_model}
\begin{figure}[!t]
\begin{center}
\includegraphics[scale=0.275]{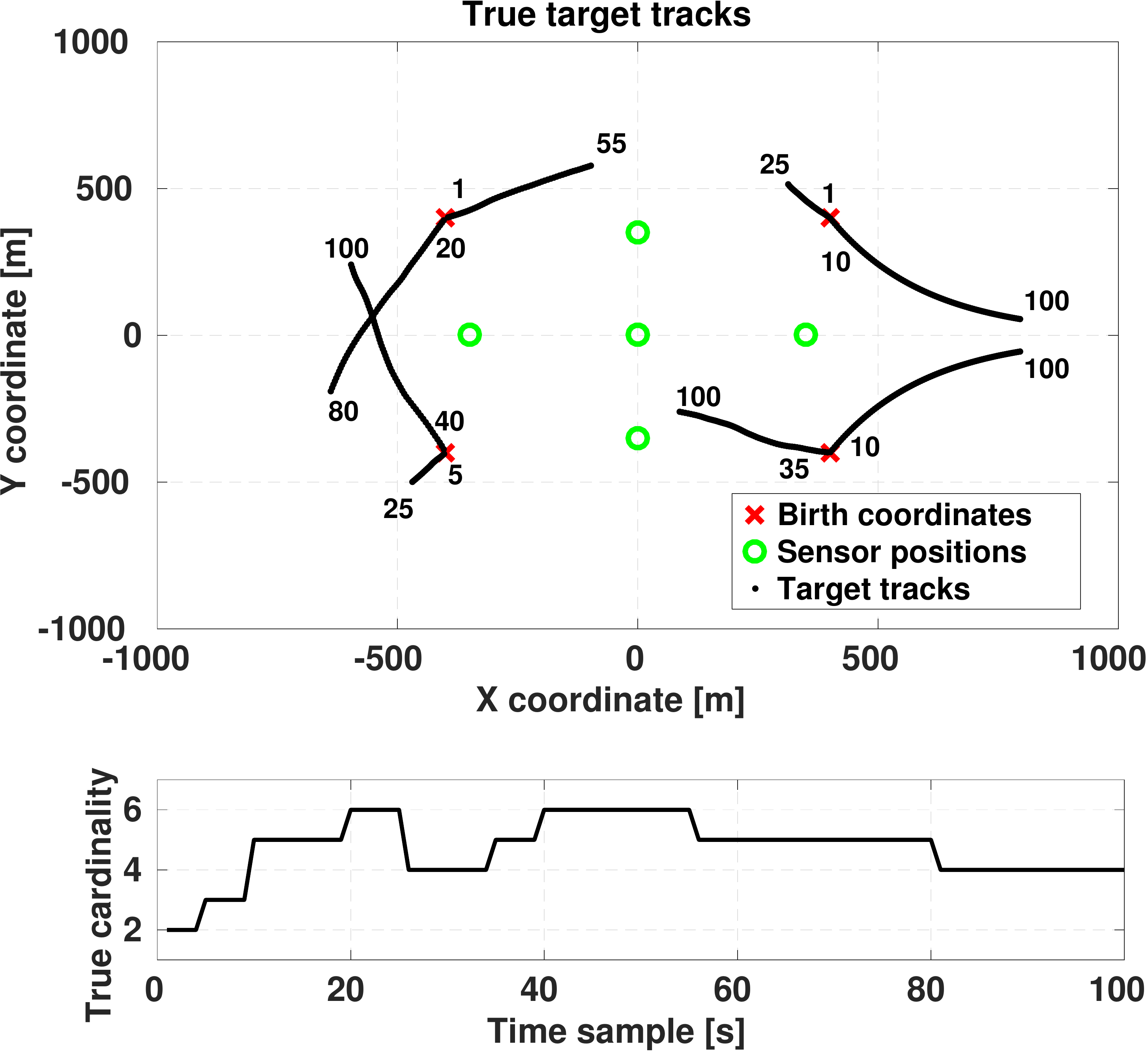}
\caption{The true number of targets and their tracks. Targets are born at locations marked with \textcolor{red}{${\times}$}. Sensor placements (only relevant in the non-linear case) are marked with \textcolor{green}{${\bigcirc}$}. }
\label{fig:true_tracks}
\end{center} 
\vspace{-3mm}
\end{figure}

Targets are assumed to evolve in a two dimensional Cartesian system. Target state vectors are taken to be $\vect{x} = [x,\, y, \, \dot{x}, \, \dot{y}]^T$, where $x$ and $y$ represent the target coordinates and $\dot{x}$ and $\dot{y}$ are its velocities along the two axes. The kinematic model for the $i$-th target is a white noise acceleration model:
\begin{equation}
\vect{x}_{k+1,i} = \mat{F}_{k+1} \vect{x}_{k,i} + \vect{v}_{k+1,i}.
\label{eq:target_kin_model}
\end{equation}
\noindent The state transition matrix is defined as
\mbox{$\mat{F}_k = \bigl [\begin{smallmatrix}
      \mat{I}_2   & T_S\, \mat{I}_2           \\
       \mat{0}_2  & \mat{I}_2          
\end{smallmatrix} \bigr]$}
where $T_s = 1$s is the sampling period; $\mat{0}_n$ and $\mat{I}_n$ are the zero and identity matrices of size $n$. The process noise is taken to be $\vect{v}_k \sim \mathcal{N}(\vect{0},\mat{Q}_k)$ with 
\mbox{$\mat{Q}_k = \sigma_{\vect{v}}^2\bigl [\begin{smallmatrix}
      \frac{T_S^3}{3}\mat{I}_2   & \frac{T_S^2}{2}\, \mat{I}_2           \\
       \frac{T_S^2}{2}\mat{I}_2  & T_S \mat{I}_2          
\end{smallmatrix} \bigr]$}.
The target tracks are depicted in Fig.~\ref{fig:true_tracks}, where a single simulation run is composed of $100$ scans sampled with $T_S = 1$s. Targets are born at locations $(\pm 400m, \pm400m)$ with time of birth and death indicated alongside their respective tracks in Fig.~\ref{fig:true_tracks}. The tracking domain is restricted to the $2000m\: \times \: 2000m$ square. The probability of survival of targets is $p_S(\vect{x}) = 0.99$ and is constant throughout the surveillance region. All filters employ a process noise of $\sigma_{\vect{v}} = 1$. In the following experiments the target tracks are kept identical throughout all Monte Carlo simulations, and the measurement noise is randomly generated at each run.

\subsection{Linear Gaussian measurement model}
\label{sec:lin_gauss_res}
In this section, targets are observed through a linear-Gaussian measurement equation and we evaluate the performance of $4$ filters: a Gaussian mixture implementation of the IC-CBMeMBer filter, a Gaussian mixture implementation of the MS-MeMBer filter as described in Sec. \ref{sec:implementation}, the Gaussian mixture implementation of the MS-CPHD filter given in \cite{bibl:nannuru_MS_CPHD_2016} and of the MS-TCPHD filter.

In this scenario, the target measurement model for the $i$-th target is 
\begin{equation}
\vect{z}_{k,i} = \mat{H}_k \vect{x}_{k,i} + \vect{w}_{k,i},
\label{eq:meas_gauss}
\end{equation}
\noindent where the observation matrix is \mbox{$\mat{H}_k = \bigl [\begin{smallmatrix}
      1 &  0 & 0 & 0           \\
      0 & 1 & 0 & 0         
\end{smallmatrix} \bigr]$}. The measurement noise is independent from the target states and is modeled as $\vect{w}_{k,i} \sim \mathcal{N}(\vect{0},\mat{R})$ with $\mat{R} = \sigma_{\vect{w}}^2 \mat{I}_2$. In our simulations, the measurement noise has $\sigma_{\vect{w}} = 10$m. Furthermore, the probability of detection of the sensors is constant throughout the surveillance region and takes the value of $p_D = 0.3$, $0.5$, $0.7$, or $0.9$. In addition to target measurements, each sensor has clutter measurements. We consider a Poisson clutter process with an average number of clutter points equal to $\lambda_c = 5$ and uniformly spread throughout the surveillance region. The clutter process is identical for all sensors.   

\begin{figure}[!t]
\begin{center}
\includegraphics[scale=0.33]{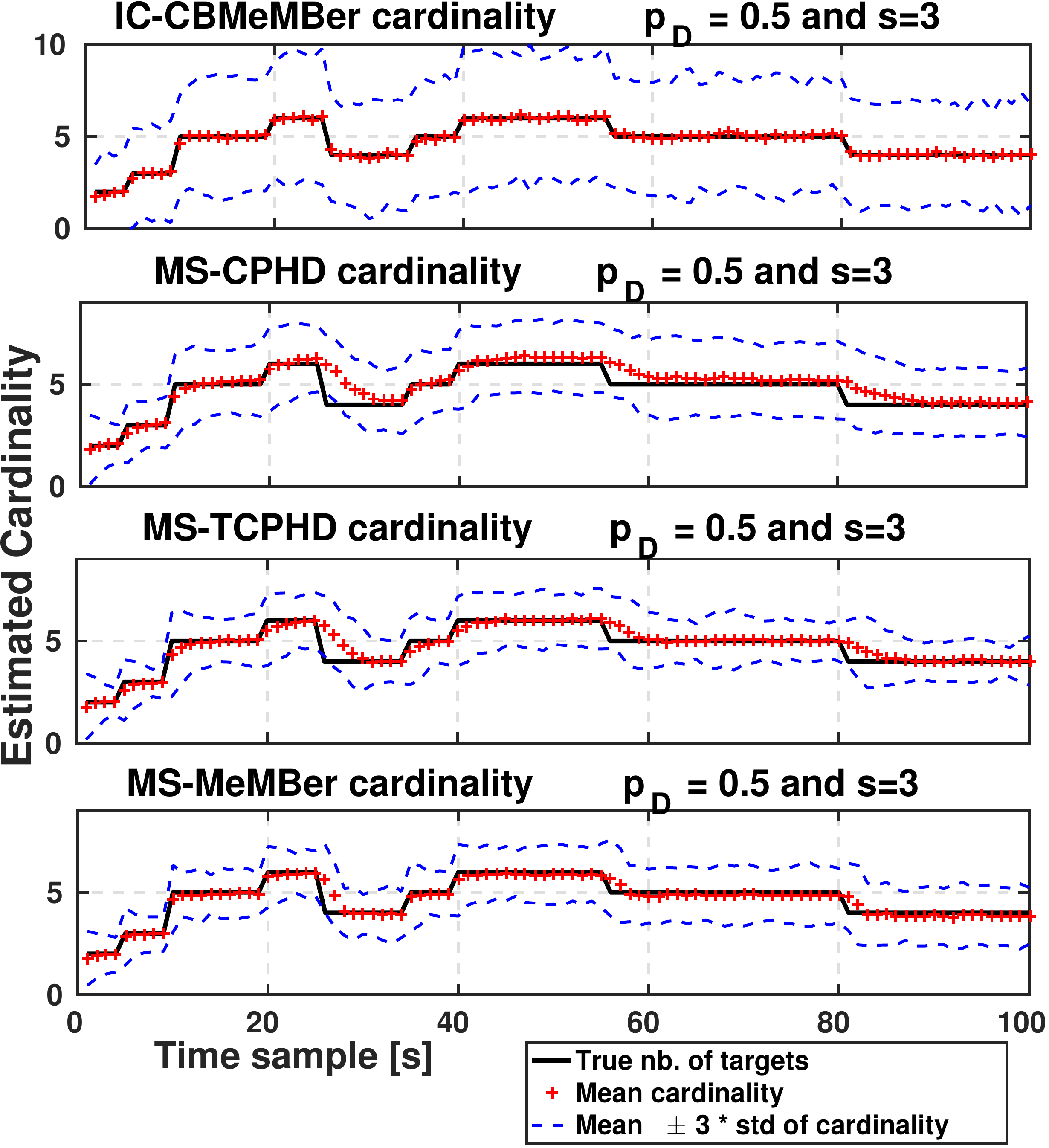}
\caption{Linear-Gaussian case: mean estimated cardinality for the IC-CBMeMBer, MS-CPHD, MS-TCPHD, and MS-MeMBer filters with $s=3$ sensors each having $p_D = 0.5$.}
\label{fig:card_est_lin_gauss}
\end{center} 
\vspace{-3mm}
\end{figure}
Gaussian mixture implementations are considered for the multi-sensor CPHD (MS-CPHD) and multi-sensor TCPHD (MS-TCPHD) filters. The MS-CPHD filter represents an exact implementation of \cite{bibl:nannuru_MS_CPHD_2016} with the PHD function represented as a Gaussian mixture. The MS-TCPHD filter also employs a Gaussian mixture PHD function. In both filters the Gaussian components are thresholded at a value of $10^{-3}$ and merging is performed subsequently.

For a fair comparison, in the Gaussian mixture IC-CBMeMBer and MS-MeMBer filters, we suppose that each Bernoulli component is represented by a single Gaussian density. For the IC-CBMeMBer filter, the Bernoulli components are pruned at a threshold of $10^{-3}$ while their number is limited at $10$ components per estimated target. For MS-MeMBer filter, the Bernoulli components are pruned at the threshold of $0.05$ and their number is limited to $4$ per estimated target. The multi-sensor subset and partition formation mechanism described in Sec. \ref{sec:implementation} is employed for the MS-MeMBer filter. 

For the MS-CPHD, MS-TCPHD and MS-MeMBer filter implementations, the maximum number of multi-sensor subsets $W_{\text{max}}$ and quasi-partitions $P_{\text{max}}$ are set to $4$, as these values were observed in \cite{bibl:nannuru_MS_CPHD_2016} and in our case to yield the best results. The birth density has $4$ components with Gaussian probability densities located at $(\pm 400m, \pm400m)$ and with identical covariance matrices $\mat{P} = \text{diag}(60, \, 60, \, 25, \, 25)$. In the MS-CPHD and MS-TCPHD filters, the birth PHD is modeled as a mixture of the aforementioned Gaussian densities each weighted with $0.1$, resulting in an average number of birthed targets of $0.4$ while the birth cardinality is Poisson distributed. In the MS-MeMBer filter, we achieve a similar birth process by supposing $4$ Bernoulli components having the same Gaussian probability density and probability of existence of $0.1$.   

In Fig.~\ref{fig:card_est_lin_gauss}, we present the mean estimated cardinality for all algorithms coupled with their respective standard deviations. For this case, we employed $s=3$ sensors all with the same probability of detection $p_D = 0.5$ while the mean values are reported over $100$ Monte Carlo runs. Observe that the MS-CPHD has a slightly higher cardinality variance than the MS-TCPHD and MS-MeMBer filters. Additionally, the cardinality standard deviation of the IC-CBMeMBer filter is significantly higher than the other filters. We noticed a faster detection of target deaths for the IC-CBMeMBer filter in comparison with the MS-CPHD, MS-TCPHD and MS-MeMBer filters. A track is terminated when the corresponding probability of existence decreases below a pre-set threshold. Besides the decrease due to survival thinning, the probability of the track is further decreased due to the update with the empty subset $W_{1:s} = \emptyset_{1:s}$. Survival thinning is identical in all filters, however the update with $\emptyset_{1:s}$ is handled differently. In addition to a decrease in component weight due to the update with $\emptyset_{1:s}$, the MS-MeMBer track probability of (\ref{eq:up_components_def1}) is also weighted with the partition score $\alpha_P$. Similarly, the PHD mixture weights of the MS-CPHD and MS-TCPHD filters are decreased due to $\emptyset_{1:s}$ but also weighted with the partition scores (see coefficient $\alpha_0$ from eq. (23) of \cite{bibl:nannuru_MS_CPHD_2016}). Note that partition scores can be high even if one subset score is low. This results in a slower track termination for the partitioning-based algorithms: MS-CPHD, MS-TCPHD and MS-MeMBer filters.

\begin{figure}[!t]
\begin{center}
    \begin{subfigure}[b]{0.5\textwidth}     \centering
        \includegraphics[scale=0.22]{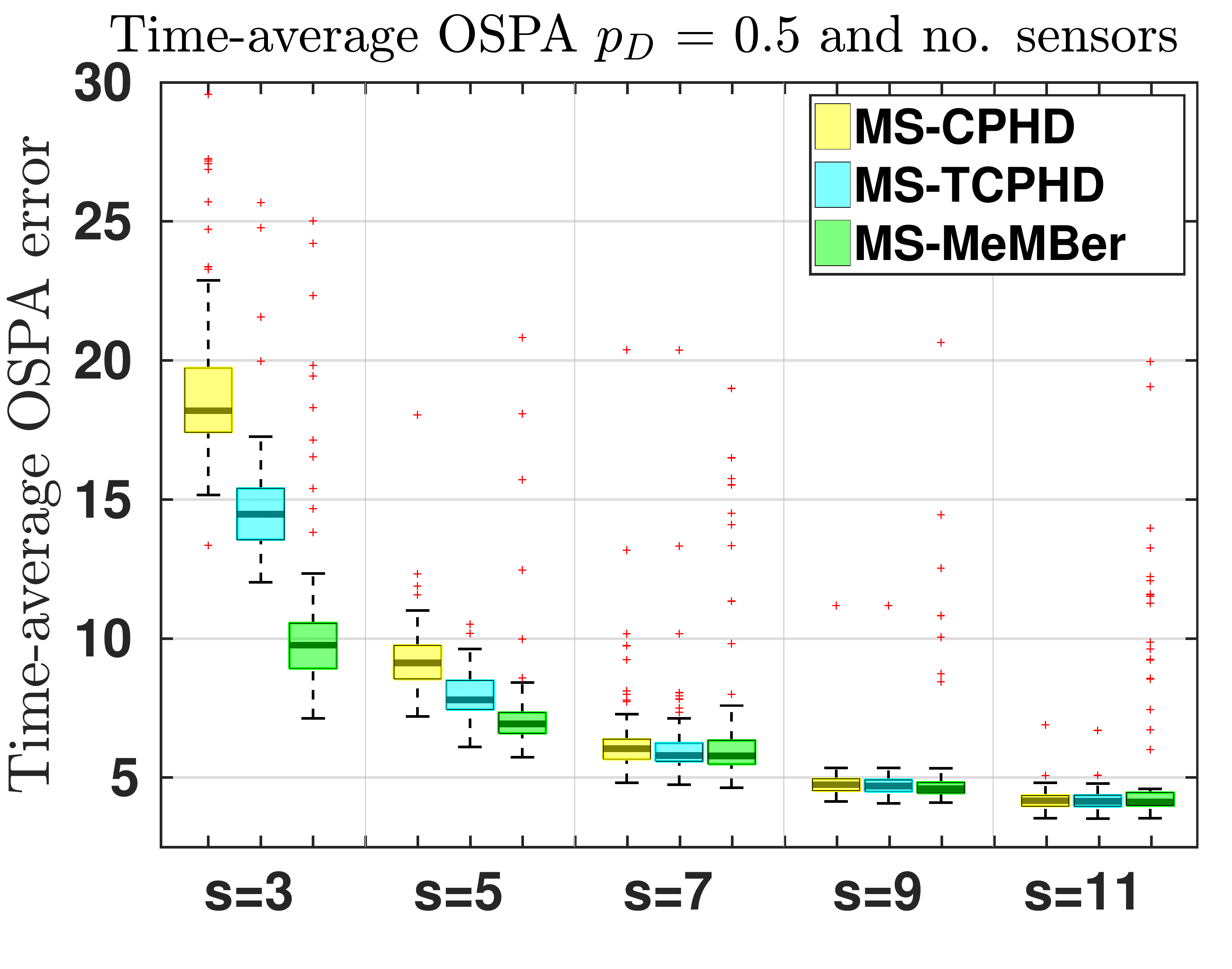}       
		\label{fig:ospa_lin_gauss_pD05}
    \end{subfigure}
     \qquad ~ 
    \begin{subfigure}[b]{0.5\textwidth}     \centering
         \includegraphics[scale=0.22]{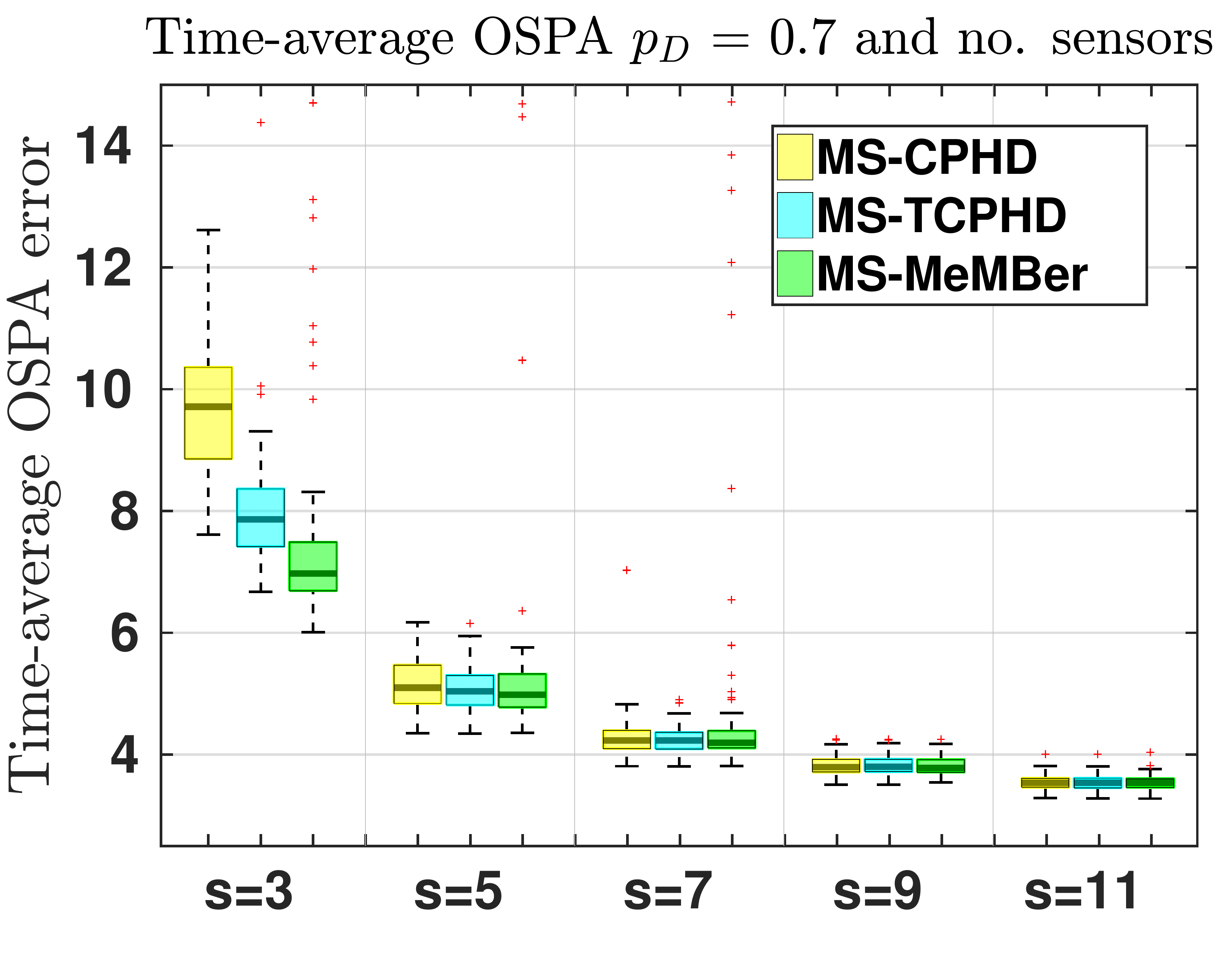}       
		 \label{fig:ospa_lin_gauss_pD07}
    \end{subfigure}
    \qquad~ 
    \begin{subfigure}[b]{0.5\textwidth}    \centering
         \includegraphics[scale=0.22]{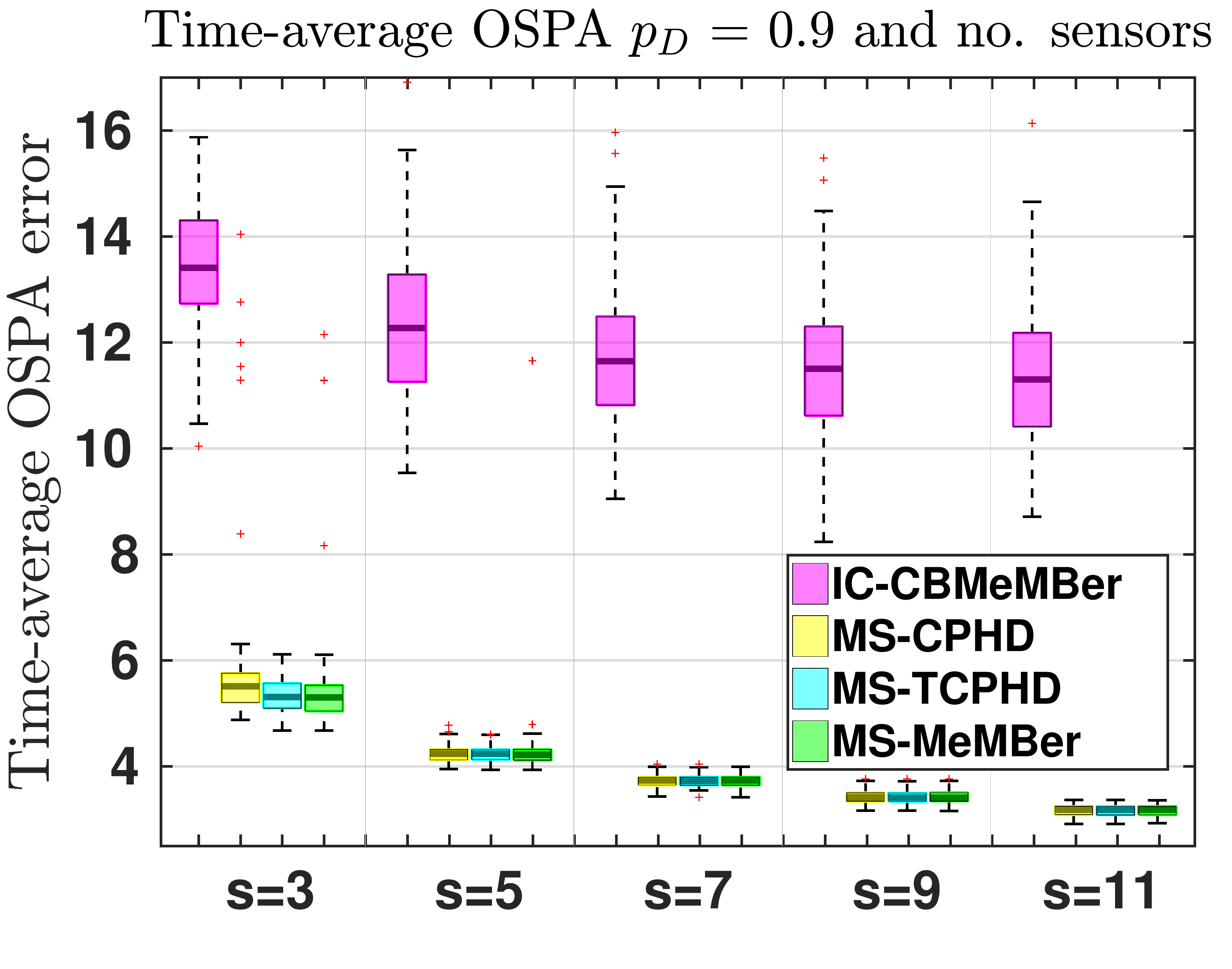}     
		 \label{fig:ospa_lin_gauss_pD09}
    \end{subfigure}
    \caption{Linear-Gaussian case: box plot of OSPA errors for different values of $p_D $ and $s$. The horizontal bar reflects the median value while the box width reflects the first and third quartiles.}\label{fig:ospa_lin_gauss}
\end{center} 
\vspace{-7mm}
\end{figure}

A comparison of time-averaged OSPA errors (that is, averaged over the $100$ scans of a single run) reported over the same $100$ runs for all filter implementations is given in Fig.~\ref{fig:ospa_lin_gauss}. These results are summarized in three box plots with $p_D = 0.5$, $p_D = 0.7$ and $p_D = 0.9$ and where each box plot showcases the OSPA errors as a function of the number of sensors $s$. Observe that OSPA errors decrease for an increased number of sensors and/or increased $p_D$. As $p_D$ and $s$ increase, the MS-CPHD, MS-TCPHD and MS-MeMBer methods converge in terms of OSPA performance. The performance of the IC-CBMeMBer filter is shown for the $p_D=0.9$ case where it is outperformed by the MS-CPHD, MS-TCPHD and MS-MeMBer methods. Additionally, the performance of IC-CBMeMBer filter was observed to decrease significantly for $p_D=0.5$ and $p_D =0.7$ since the filter derivation is based on the assumption of high $p_D$. In the $p_D=0.9$ case, note the relative slowly increase in performance of the IC-CBMeMBer filter with increasing number of sensors $s$. The IC-CBMeMBer filter sequentially applies the CBMeMBer update step for each sensor which leads to the accumulation of approximation errors due to the assumptions of high $p_D$, low density clutter and first-order moment approximations applied at each update step. In contrast, the MS-CPHD, MS-TCPHD and MS-MeMBer methods simultaneously use the measurements from all sensors (under the form of multi-sensor subsets) to obtain the exact updated posterior which is subsequently approximated as an iid cluster or a multi-Bernoulli RFS respectively. The MS-CPHD and MS-TCPHD filters are similar with the difference being in the associations between the predicted PHD components and multi-sensor subsets. Although the MS-MeMBer filter propagates the full posterior density and employs a different RFS model than the CPHD filters, it also resorts to a first-order approximation (i.e., a PHD approximation) in the derivation of its updated posterior. Additionally, notice the increased outliers in the box plots of Fig.~\ref{fig:ospa_lin_gauss} for lower $p_D$, which are generated when the filters exhibit a high number of cardinality errors. Such errors occur more often for smaller $p_D$ values and/or smaller number of sensors.

Furthermore, observe the improved performance of the MS-MeMBer filter algorithm at low $p_D$ and/or reduced number of sensors. Reducing $p_D$ and/or $s$ leads to an increase of the weights of the miss-detected (i.e., legacy) components of the updated PHD function of the MS-CPHD and MS-TCPHD filters. On subsequent times, the update step of the MS-CPHD filter forms all associations between the PHD mixture components (including the previous-time legacy components) and the measurement subsets within a partition. In the case of the MS-CPHD filter, this translates to higher weights for the previous-time legacy components as compared to the MS-TCPHD filter. As a result, the MS-CPHD filter has a more dispersed cardinality distribution estimate and hence a higher OSPA error. The MS-TCPHD and MS-MeMBer filters correct this by updating each component with its best-scoring multi-sensor subset as given by the greedy method for subsets of Section \ref{sec:subset_form}. The difference between the MS-MeMBer filter and MS-TCPHD filter resides in the specific form of the probabilities of existence and component weights which are intrinsic to their prior distributions, i.e., multi-Bernoulli and iid cluster respectively. Numerically, at lower values of $s$ and $p_D$ the mixture weights of the MS-CPHD filter were found to be dispersed between $[0,1]$ whereas the probabilities of existence of the MS-MeMBer filter were found to be more concentrated around the limits $0$ and $1$, which explains the improved cardinality estimates of the MS-MeMBer filter in Fig.~\ref{fig:card_est_lin_gauss}. 
\begin{table}[!t]
 \caption{Linear Gaussian case with $p_D = \lbrace 0.5, 0.9 \rbrace$: average computational time.}
 \label{table:linear_time_pD09}
\begin{center}
\renewcommand{\arraystretch}{1.5}
\begin{tabular}{|c|c|c|c|c|c|c|} 
\cline{1-7}

\multicolumn{1}{ |c| }{$p_D$ } & \multicolumn{1}{ c| }{ Filter} &  \multicolumn{1}{ c| }{$s=3$} & \multicolumn{1}{ c| }{ $s=5$} &  \multicolumn{1}{ c| }{$s=7$} &  \multicolumn{1}{ c| }{$s=9$} &  \multicolumn{1}{ c| }{$s=11$}\\ \cline{1-7} 
\hhline{|-|-|-|-|-|-|}

\multirow{4}{*}{$0.9$} &
{$\substack{IC-CB\\ MeMBer}$}  & $126$ms & $283$ms & $510$ms & $800$ms & $1112$ms \\  \cline{2-7}

& {$\substack{MS \\ CPHD}$}  & $39$ms & $62$ms & $86$ms & $112$ms & $133$ms \\  \cline{2-7}

& {$\substack{MS \\ TCPHD}$}  & $12$ms & $18$ms & $24$ms & $30$ms & $36$ms \\  \cline{2-7}
& {$\substack{MS \\ MeMBer}$}  & $10$ms & $16$ms & $22$ms & $29$ms & $35$ms \\  \cline{2-7}

\hhline{|=|=|=|=|=|=|=|}

\multirow{4}{*}{$0.5$} &
{{\centering$ \substack{IC-CB \\  MeMBer}$}}  & $169$ms & $303$ms & $450$ms & $681$ms & $952$ms \\  \cline{2-7}

& {$\substack{MS \\ CPHD}$}  & $47.5$ms & $67.6$ms & $71$ms & $76$ms & $88$ms \\  \cline{2-7}

& {$\substack{MS \\ TCPHD}$}  & $19.2$ms & $26.2$ms & $27$ms & $29$ms & $34$ms \\  \cline{2-7}
& {$\substack{MS \\ MeMBer}$}  & $9.7$ms & $15.2$ms & $21$ms & $27$ms & $33$ms \\  \cline{2-7}

  \hhline{|=|=|=|=|=|=|=|}
\end{tabular} 
\end{center}
\end{table}

%
%
%
%
Table~\ref{table:linear_time_pD09} summarizes the average computation time of the filters for $p_D=\lbrace 0.5, 0.9\rbrace$ and a varying number of sensors. The average computational time for a given filter is defined as the duration the filter takes to process all scans divided by the number of scans while the values reported in Table~\ref{table:linear_time_pD09} were also averaged over $100$ Monte Carlo runs. Observe that the MS-CPHD, MS-TCPHD and MS-MeMBer filters exhibit a linear complexity with respect to the number of sensors. This fact is supported by the linear complexity of the greedy subset and quasi-partition formation algorithms \ref{sec:implementation}. Additionally, observe the increased computational load of the MS-CPHD filter, which is caused by the more involved update stage, which updates each mixture component of the PHD function with all of the multi-sensor subsets from a given partition. The computational requirements of the MS-MeMBer and MS-TCPHD filters are similar. Note also the high computational requirements of the IC-CBMeMBer filter due to the high number of Bernoulli components needed to produce satisfactory results. 
\subsection{Non-linear measurement model}
\label{sec:non_lin_res}

A Doppler-bearing measurement model is considered in this section, with the measurement vector consisting of a noisy bearing and Doppler shift. The sensor coordinates are denoted by $\lbrace (x^j,y^j) \vert j=1,\dots, s\rbrace$. The measurement of the $i$-th target with state vector $\vect{x}_{i,k} = [x \; y \;\dot{x}\; \dot{y}]^T$ and collected at the $j$-th sensor is  
\begin{equation}
\vect{z}_{k,i}^j = 
\begin{bmatrix}
    \text{atan2}(\frac{y-y^j}{x-x^j})       \\
    \frac{2f_c}{c} \frac{(x-x^j)\dot{x} + (y-y^j)\dot{y}}{\sqrt{(x-x^j)^2 + (y-y^j)^2}}      
\end{bmatrix}
 + \vect{w}_{k,i}^j,
\label{eq:meas_doppler}
\end{equation}
\noindent where $\text{atan2}(\cdot)$ is the four-quadrant inverse tangent function; $f_c$ is the carrier frequency of the received signal; and $c$ is the wave velocity. In our simulations $f_c = 300$Hz and $c=1450$m/s, corresponding to an underwater scenario. The measurement noise is independent of the target states and is taken to be $\vect{w}_{k,i} \sim \mathcal{N}(\vect{0}, \mat{R})$ with $\mat{R} = \text{diag}(\sigma_{\theta}^2,\, \sigma_{f}^2)$. The bearing standard deviation is $\sigma_{\theta} = 1$ degree while the Doppler standard deviation is $\sigma_{f} = {0.7}$ Hz. Poisson distributed clutter is appended to the measurement set of each sensor. Unless otherwise stated, the average clutter rate is fixed to $5$ points per scan and the clutter density is uniform over the observation domain of $2\pi \times [- 100, + 100]$. The target tracks and kinematics are as shown in Fig.~\ref{fig:true_tracks}.

In this non-linear scenario, both UKF (Unscented Kalman Filter) and SMC implementations are considered for the MS-MeMBer, MS-TCPHD, MS-CPHD  and IC-CBMeMBer filters. The UKF filters are implemented using Gaussian mixtures in conjunction with the unscented transform \cite{bibl:julier_uhlmann_UKF2004} to achieve the non-linear measurement updates. The resulting implementations are referred to as the UKF MS-CPHD, UKF MS-TCPHD, UKF MS-MeMBer and UKF IC-CBMeMBer filters. 
\begin{figure}[!t]
\begin{center}
\includegraphics[scale=0.35]{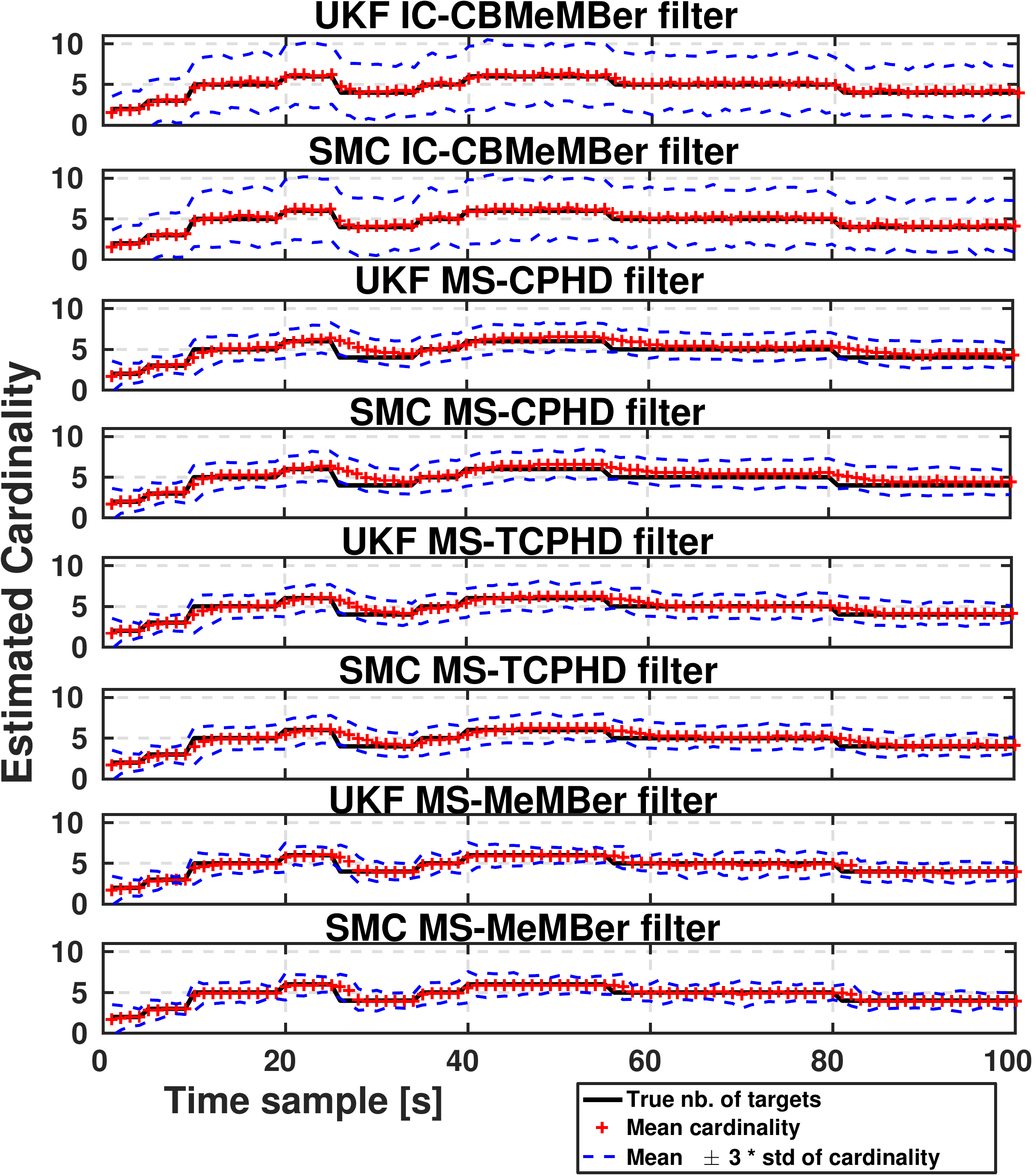}
\caption{Non-linear case: estimated cardinality of various filters for $p_D=0.3$.}
\label{fig:card_est_pf}
\end{center} 
\vspace{-3mm}
\end{figure}
In order to reduce the computational load of the greedy subset and partition selection algorithms for the SMC implementations of the MS-CPHD and MS-TCPHD filters, clustering of the predicted PHD function could be performed. In our simulation, we employ a particle PHD function as a mixture of target densities, each having a separate particle representation. To achieve this, the birth PHD function is represented as a mixture of separate particle sets. This leads to an \textit{implicitly} clustered PHD function, i.e., we avoid the use of clustering methods such as k-means. The birth PHD function is approximated by $4$ distinct particle sets centered around the birth locations $(\pm 400, \: \pm 400)$ and sampled from Gaussian probability densities with covariance matrices $\mat{P} = \text{diag}(40, \, 40, \, 25, \, 25)$. The subsequent predicted and updated PHD functions are represented as a mixture of separate particle sets, each representing a potential target. The greedy multi-sensor subset mechanism of \cite{bibl:nannuru_MS_CPHD_2016} is applied to each particle set of the predicted PHD function. Subsequently, partitions are formed from the resulting subsets.

In the case of the SMC MS-MeMBer and SMC IC-CBMeMBer filters, each Bernoulli component has a probability density represented as a set of discrete points (\ref{eq:pf_pdf}). The scoring of measurement collections $W_{1:s}$ in the SMC MS-MeMBer filter is done via (\ref{eq:pf_score}). For both filters, the birth process is composed of $4$ Bernoulli components placed at the locations $(\pm 400, \: \pm 400)$ and having covariances equal to $\mat{P} = \text{diag}(40, \, 40, \, 25, \, 25)$. 

In all SMC filter implementations, $700$ particles per target are used, and sampling is done directly from their respective birth densities or the prediction kernel. 
\begin{figure}[!t]
\begin{center}
    \begin{subfigure}[b]{0.5\textwidth}
        \includegraphics[scale=0.27]{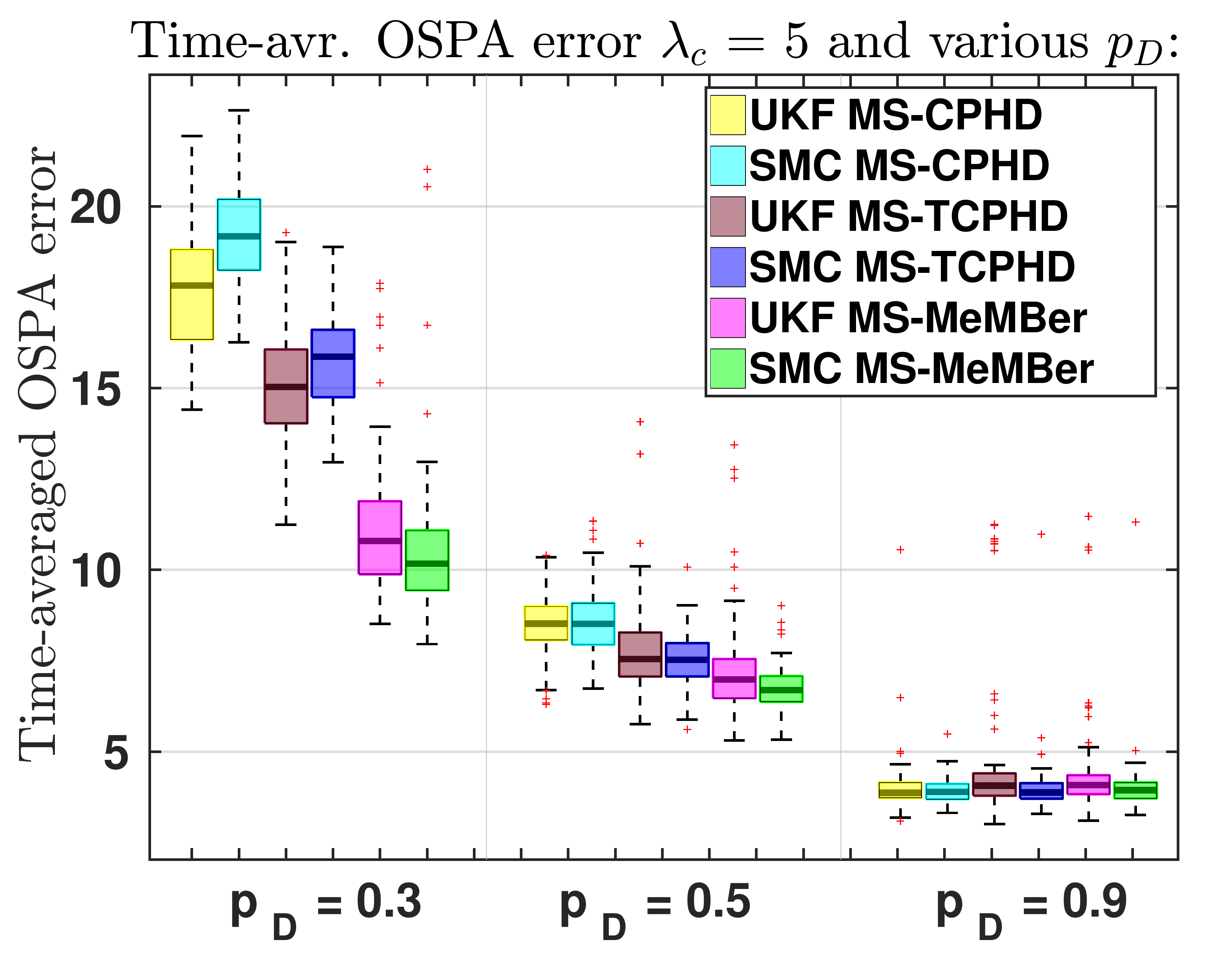}       
		\label{fig:ospa_nonlin_lc5}
    \end{subfigure}  
      \quad ~ 
    \begin{subfigure}[b]{0.5\textwidth}
         \includegraphics[scale=0.27]{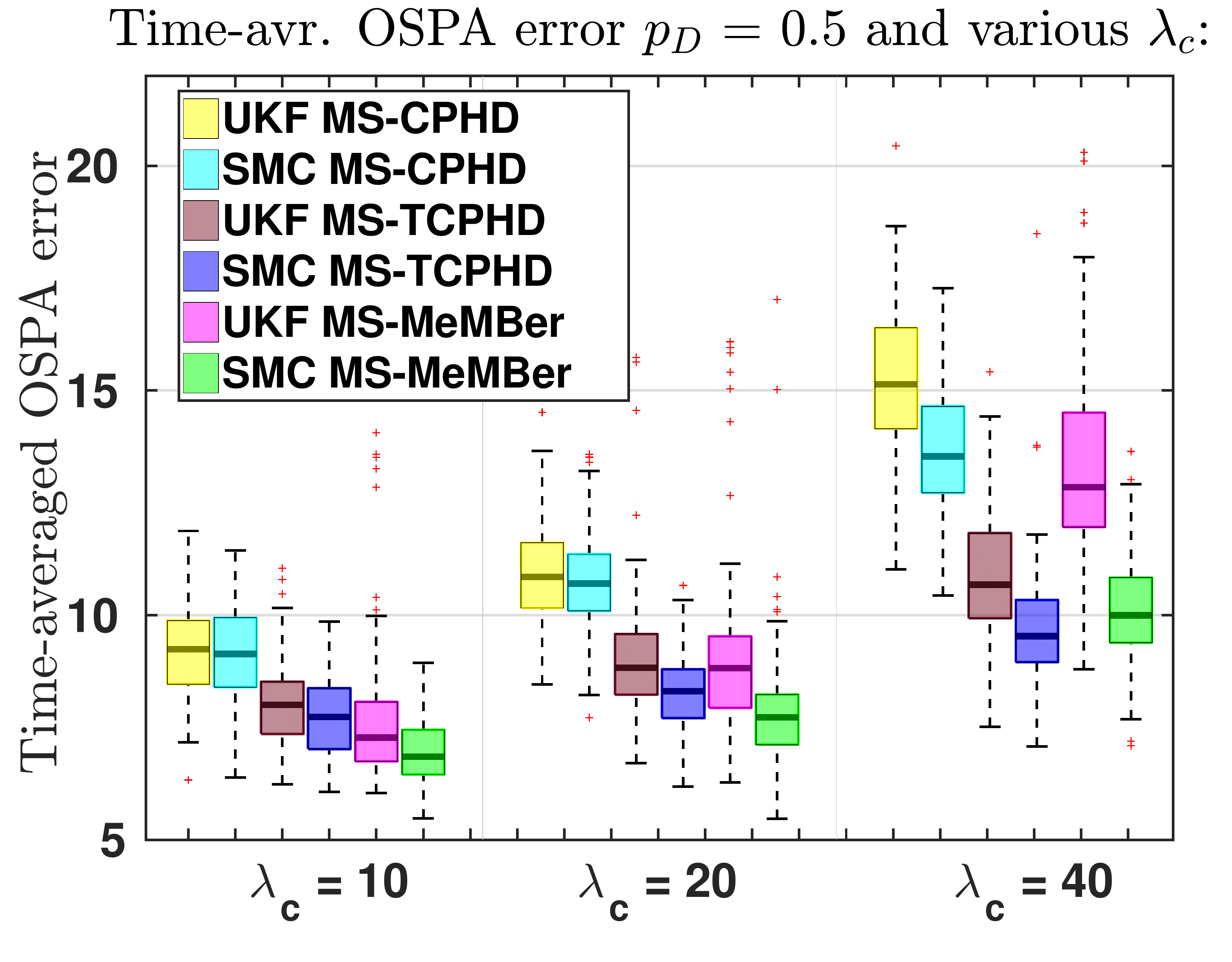}       
		 \label{fig:ospa_nonlin_lc20}
    \end{subfigure}
    \caption{Non-linear case: box plot of OSPA errors for diferent values of $p_D $ and clutter rate $\lambda_c$. The horizontal bar reflects the median value while the box width reflects the first and third quartiles.}\label{fig:ospa_non_lin}
\end{center} 
\vspace{-7mm}
\end{figure}

In this difficult non-linear tracking scenario we employ a threshold of $10^{-10}$ for the pruning of Bernoulli and Gaussian mixture components in the aforementioned filter implementations. Additionally, the number of Bernoulli components in the MS-MeMBer filter implementations is capped at $4$ per estimated target, while this value is increased to $20$ for the IC-CBMeMBer filter implementations; with fewer Bernoulli components per target, the performance of the IC-CBMeMBer filter is not comparable to the other filters, even at high $p_D$. The MS-CPHD and MS-TCPHD filter implementations cap the number of PHD mixture components per target at $4$. These values are selected on the basis of balancing tracking performance with computational overhead.    
\begin{table}[!t]
 \caption{Non-linear case: median time-averaged OSPA values, lower $Q_1$ and upper $Q_3$ quartiles are given in brackets $[Q_1,\: Q_3]$. Clutter rate is fixed to $\lambda_c=5$ per sensor.}
 \label{table:non_linear_ospa}
\begin{center}
\begin{tabular}{cc|c|c|} 
\cline{2-4}

&  \multicolumn{1}{ |c| }{$p_D=.3$} & \multicolumn{1}{ c| }{ $p_D=.5$} &  \multicolumn{1}{ c| }{$p_D=.9$}  \\ \cline{1-4} 
\hhline{|=|=|=|=|}

\multicolumn{1}{ |c| }{{\multirow{2}{*}{$\substack{SMC \\ MS-MeMBer}$}}} & 
  $10.2 $ & $6.7$ & $3.9$ \\ 
\multicolumn{1}{ |c| }{} & $[9.4,\: 11.1] $ & $[6.4,\: 7.1]$ & $[3.7,\: 4.1]$ \\  \cline{1-4}

\multicolumn{1}{ |c| }{{\multirow{2}{*}{$\substack{UKF \\ MS-MeMBer}$}}} & 
  $10.8$ & $6.9$ & $4.1$ \\ 
\multicolumn{1}{ |c| }{} & $[9.9,\: 11.9] $ & $[6.5,\: 7.5]$ & $[3.8,\: 4.3]$ \\  \cline{1-4}

\multicolumn{1}{ |c| }{{\multirow{2}{*}{$\substack{SMC \\ MS-CPHD}$}}} & 
  $19.2$ & $8.5$ & $3.9$ \\ 
\multicolumn{1}{ |c| }{} & $[18.2,\: 20.2] $ & $[7.9,\: 9.1]$ & $[3.7,\: 4.1]$ \\  \cline{1-4}

\multicolumn{1}{ |c| }{{\multirow{2}{*}{$\substack{UKF \\ MS-CPHD}$}}} & 
  $17.8$ & $8.5$ & $3.9$ \\ 
\multicolumn{1}{ |c| }{} & $[16.3,\: 18.8] $ & $[8.1,\: 9]$ & $[3.7,\: 4.2]$ \\  \cline{1-4}

\multicolumn{1}{ |c| }{{\multirow{2}{*}{$\substack{SMC \\ MS-TCPHD}$}}} & 
  $15.8$ & $7.5$ & $3.9$ \\ 
\multicolumn{1}{ |c| }{} & $[14.7,\: 16.6] $ & $[7.1,\: 8]$ & $[3.7,\: 4.1]$ \\  \cline{1-4}

\multicolumn{1}{ |c| }{{\multirow{2}{*}{$\substack{UKF \\ MS-TCPHD}$}}} & 
  $15$ & $7.5$ & $4.1$ \\ 
\multicolumn{1}{ |c| }{} & $[14,\: 16.1] $ & $[7.1,\: 8.3]$ & $[3.8,\: 4.4]$ \\  \cline{1-4}

\multicolumn{1}{ |c| }{{\multirow{2}{*}{$\substack{SMC \\ IC-CBMeMBer}$}}} & 
  $36.4$ & $31.9$ & $13.1$ \\ 
\multicolumn{1}{ |c| }{} & $[35.6,\: 37.7] $ & $[31.1,\: 32.9]$ & $[12.2,\: 13.9]$ \\  \cline{1-4}

\multicolumn{1}{ |c| }{{\multirow{2}{*}{$\substack{UKF \\ IC-CBMeMBer}$}}} & 
  $36.1$ & $31$ & $12$ \\ 
\multicolumn{1}{ |c| }{} & $[35.1,\: 37] $ & $[30.2,\: 32.1]$ & $[11.4,\: 13.1]$ \\  \cline{1-4}
  \hhline{|=|=|=|=|}
\end{tabular} 
\end{center}
\end{table}
We consider the following sensor configuration with $s=5$ sensors placed at coordinates $(x,y) \in \lbrace  (-350,0),\;(350,0),\;(0,0),\;(0,-350),\;(0,350)  \rbrace$ as seen in Fig.~\ref{fig:true_tracks}. Throughout the following simulations, the sensors have equal probabilities of detection that are constant over the surveillance region. Note that the SMC implementations are capable of handling non-constant $p_D(\vect{x})$, as opposed to the UKF implementations. 

For $100$ Monte Carlo simulations, the mean and standard deviation of the cardinality estimates of the various filters are shown in Fig.~\ref{fig:card_est_pf} for the case of $p_D = 0.3$. Notice the overall improved cardinality estimate of the MS-MeMBer filter and especially of the SMC MS-MeMBer. Furthermore, notice the poor performance of both implementations of the IC-CBMeMBer filter. We observed that the performance of the IC-CBMeMBer filter only becomes comparable to the performance of the MS-MeMBer filter when all sensors have very high probabilities of detection, e.g, $0.98$. Indeed, the IC-CBMeMBer filter relies on the application of the CBMeMBer measurement correction step (and the ensuing update approximations) sequentially for each sensor. Hence, the IC-CBMeMBer filter requires a high detection probability at each sensor, whereas the generalized multi-sensor variants of the CPHD and MeMBer filters perform a simultaneous update step with all sensor measurements before resorting to any additional approximations.
\begin{figure}[!t]
\begin{center}
\includegraphics[scale=0.27]{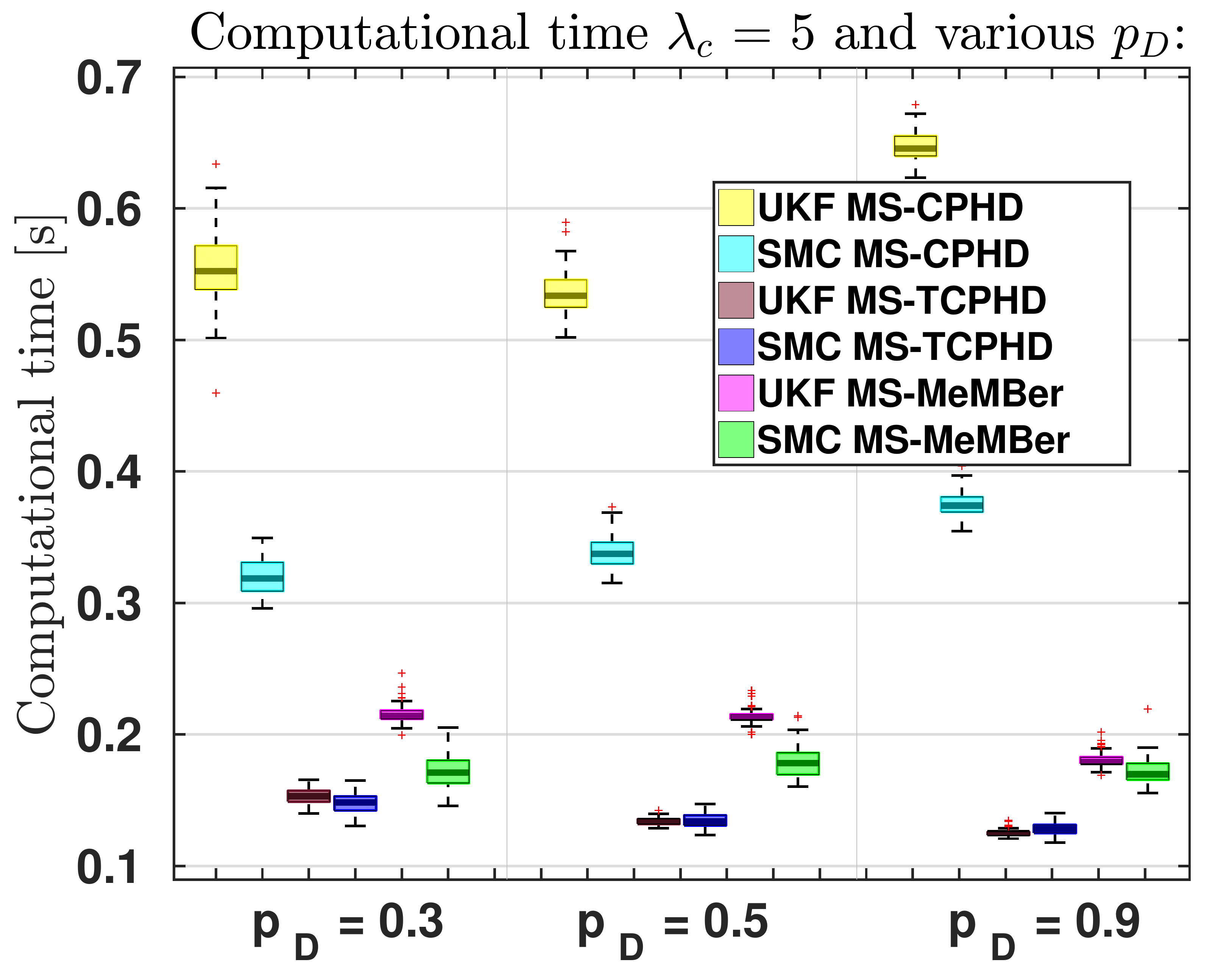}
\caption{Non-linear case: average computation times for various filter implementations at different probabilities of detection. The horizontal bar reflects the median value while the box width reflects the first and third quartiles.}
\label{fig:box_comp_times_S1}
\end{center} 
\vspace{-3mm}
\end{figure}
The time-averaged OSPA errors (i.e., averaged over the $100$ scans of a single run) from the $100$ Monte Carlo simulations, are displayed in Fig.~\ref{fig:ospa_non_lin}. The figure shows how the errors change as we vary the probability of detection or the average number of clutter points per sensor. Notice the improved performance of the MS-MeMBer filter for lower $p_D$. In the case of the MS-CPHD filters at lower $p_D$, the update step of a specific predicted component includes significant contributions from subsets that correlate highly with other components and which degrade its performance. The UKF and SMC MS-MeMBer filters both provide better cardinality estimates as compared with their CPHD counterparts. This is attributed to the different RFS models, i.e., the iid cluster and the multi-Bernoulli RFS, employed by the two filters in this simulation scenario. As in the linear case, for low $p_D$ the normalized mixture weights of the MS-CPHD filter were found to be dispersed between $[0,1]$ whereas the probabilities of existence of the MS-MeMBer filter were found to be more concentrated around the limits $0$ and $1$. This difference between the concentration of weights and probabilities of existence leads to a difference in their estimated cardinality distributions. However, at very high clutter rates ($\lambda_c = 40$ per sensor), the SMC MS-TCPHD filter and the SMC MS-MeMBer filter have comparable performance. As in the linear-Gaussian case, for high $p_D$ all methods converge in terms of OSPA performance.

In Table \ref{table:non_linear_ospa} we present several filtering results for $p_D \in \lbrace 0.3,\, 0.5, \,  0.9\rbrace$. Time-averaged OSPA errors are recorded for each of the $100$ Monte Carlo simulations, while the median, the lower and the upper quartile are shown in Table \ref{table:non_linear_ospa}. Notice again the convergence of filters in terms of OSPA error as $p_D$ increases and a significant advantage of the MS-MeMBer filters at lower $p_D$ values. Both implementation of the MS-CPHD and MS-TCPHD filters struggle at $p_D = 0.3$ due to increased cardinality errors. Notice also the poor performance of the UKF and SMC IC-CBMeMBer filters for all values of $p_D$.  

Computation times are shown in the box plot of Fig.~\ref{fig:box_comp_times_S1} for different $p_D$ values. The average computational time for a given filter represents the duration the filter takes to process all scans divided by the number of scans. 

Compared to the MS-TCPHD and MS-MeMBer filters, the MS-CPHD filter is the most computationally expensive because each multi-sensor subset identified by the greedy subset identification algorithm is used to update {\em all} predicted Gaussians or particle groups. In contrast, the MS-TCPHD and MS-MeMBer filters only update the single predicted component that best matches each multi-sensor subset (as measured by the score function). In this simulation, the UKF updates are more expensive than the SMC updates because they involve multiple matrix inversions. Additionally, observe an increase of computational requirements as $p_D$ increases. An increase of $p_D$ leads to an increased average number of measurements per sensor which in turn increases the computational cost of the greedy multi-sensor subset selection method.

\section{Conclusions}
\label{sec:conclusions}
In this paper a multi-sensor multi-Bernoulli filter is derived for multi-target tracking. The proposed filter partitions the multi-sensor observations into multi-sensor subsets which are associated with the Bernoulli components. We describe computationally tractable approximate Gaussian mixture and particle filter implementations. The filter is shown to have a reduced computational load compared to the current implementations of the multi-sensor CPHD filter and improved performance at low probability of detection. 


%

\appendices
\section{Proof of Lemma \ref{the:derivated_funcl}}
\label{app:proof_deriv_func}

In this appendix, the time index is dropped to simplify notation. Subscripts refer to sensors and superscripts refer to Bernoulli components. For example, $W_i^j$ denotes the subset of measurements from sensor $i$ associated with the PGFl of the $j$-th Bernoulli component while $W_i^0$ denotes the clutter subset from sensor $i$. Note that $  W^0_i, W^1_i, \cdots, W^M_i  $ form a quasi-partition of the set of measurements $Z_i$, i.e., the subsets are allowed to be empty and $ \uplus_{j=0}^M W^j_i =  Z_{i}$ with $\uplus$ indicating the disjoint union operator. The condition of at most one measurement per target per sensor translates to having $\card{W^j_i} \leq 1$ for $j =1,\dots, M$ and $i =1,\dots, s$. Furthermore, whenever $W^j_i= \emptyset$, the corresponding differential becomes $\frac{\delta G^j}{\delta W^j_i}[g] \triangleq G^j[g]$ \cite[Eq. 11.191]{bibl:mahler_book2007}.

For an arbitrary number of sensors $k$, we denote the ordered collections $W^j_{1:k} = ( W^j_1, \cdots, W^j_k )$ for $j =0,\dots, M$ and by a slight abuse of notation we introduce $\frac{\delta G^j}{\delta W_{1:k}^j} \triangleq \frac{\delta^k G^j}{\delta W_{1}^j \cdots \delta W_{k}^j}$, where each differential $\frac{\delta}{\delta W_i^j}$ is taken in $g_i(\vect{z})$ with respect to the measurement set $W_i^j$. Let $P_{1:k} = ( W^0_{1:k}, W^1_{1:k}, \cdots, W^M_{1:k}  )$ and $\mathcal{P}_{1:k}$ be the set of all collections $P_{1:k}$ that respect $ \uplus_{j=0}^M W^j_i =  Z_{i}$ for each sensor $i=1, \dots,k$ and $\card{W^j_i} \leq 1$ for $j =1,\dots, M$ and $i =1,\dots, s$.

The proof of Lemma \ref{the:derivated_funcl} involves differentiating the functional (\ref{eq:up_func_dev2}) $s$ times (i.e., with respect to sensors $1, \dots, s$) and is carried out in two different stages. Note that (\ref{eq:up_func_dev2}) is a product of Bernoulli PGFls and clutter pgfs. In a first step, we employ mathematical induction and the general product rule \cite[Eq. 11.274]{bibl:mahler_book2007} in order to write the differentiation of (\ref{eq:up_func_dev2}) for an arbitrary $s$ as an expression involving the Bernoulli and clutter derivatives. The differentiation of the individual Bernoulli PGFls and clutter pgfs is carried out in a second step.

\textbf{Induction base $k=1$}. The differentiation of the functional (\ref{eq:up_func_dev2}), via the general product rule \cite[Eq. 11.274]{bibl:mahler_book2007}, for the first sensor is given by
\begin{equation}
\frac{\delta F}{\delta Z_1}[g_{1:s},u] = \left[ \prod_{i=2}^s C_i(\dotprod{c_i}{g_i}) \right] 
 \sum_{W^0_1 \uplus W_1^1 \uplus \cdots \uplus W_{1}^M =  Z_{1} } \frac{\delta C_1}{\delta W^0_1} \frac{\delta G^1}{\delta W^1_1} \cdots \dfrac{\delta G^M}{\delta W^M_1}.
\label{eq:k1_diffA}
\end{equation}
Note that the sum of (\ref{eq:k1_diffA}) comprises additional terms corresponding to partitions of the set $\set{Z}_1$ that do not respect the at most one measurement per target condition. However, these terms vanish since the functionals $G^j[u \textstyle{\prod_{i=1}^s \phi_{g_i}}]$ for $j=1, \dots, M$ are linear with respect to the function $g_1(\cdot)$. Employing the $P_{1:k}$ and $\mathcal{P}_{1:k}$ notation for $k=1$, we can rewrite (\ref{eq:k1_diffA}) as
\begin{equation}
\frac{\delta F}{\delta Z_1}[g_{1:s},u] = \left[ \prod_{i=2}^s C_i(\dotprod{c_i}{g_i}) \right] 
 \sum_{\substack{P_1 \in \mathcal{P}_1 \\ P_1 = ( W_1^0, W_1^1, \cdots, W_1^M)}} \frac{\delta C_1}{\delta W^0_1} \frac{\delta G^1}{\delta W^1_1} \cdots \dfrac{\delta G^M}{\delta W^M_1} .
\label{eq:k1_diff_genA}
\end{equation}
\textbf{Induction step for $k$ with $k < s-1$}. Consider that the differentiation of the functional (\ref{eq:up_func_dev2}) with respect to the first $k$ sensors has the form
\begin{equation}
\frac{\delta^k F}{\delta Z_k \cdots \delta Z_{1}}  = \left[ \prod_{i=k+1}^s C_i(\dotprod{c_i}{g_i}) \right] 
 \sum_{\substack{P_{1:k} \in \mathcal{P}_{1:k} \\ P_{1:k} = ( W_{1:k}^0, W_{1:k}^1, \cdots, W_{1:k}^M)}} \left[ \prod_{i=1}^k\frac{\delta C_i}{\delta W^0_i} \right] \frac{\delta G^1}{\delta W^1_{1:k}} \cdots \dfrac{\delta G^M}{\delta W^M_{1:k}} .
\label{eq:kk_diff_genA}
\end{equation}
We are interested in the differentiation of (\ref{eq:kk_diff_genA}) in $g_{k+1}(\vect{z})$ with respect to the measurements of sensor $Z_{k+1}$, i.e.,
\begin{align}
&\frac{\delta}{\delta Z_{k+1}} \left\lbrace \frac{\delta^k F}{\delta Z_k \cdots \delta Z_{1}}  \right\rbrace = \left[ \prod_{i=k+2}^s C_i(\dotprod{c_i}{g_i}) \right] \times \nonumber \\
&   \sum_{W^0_{k+1} \uplus W_{k+1}^1 \uplus \cdots \uplus W_{k+1}^M =  Z_{k+1} } \sum_{P_{1:k} \in \mathcal{P}_{1:k}} \left[ \left( \prod_{i=1}^{k+1}\frac{\delta C_i}{\delta W^0_i} \right)   \frac{\delta G^1}{\delta W^1_{k+1} \delta W^1_{1:k}} \cdots \dfrac{\delta G^M}{\delta W^M_{k+1} \delta W^M_{1:k}} \right] ,
\label{eq:kk1_diffA}
\end{align}
where the partitioning of the measurement set $ Z_{k+1} = W^0_{k+1} \uplus W_{k+1}^1 \uplus \cdots \uplus W_{k+1}^M  $ is given by the general product rule. Introducing $W_{1:k+1}^j = \left( W_{1:k}^j, \, W_{k+1}^j \right)$ for $j=0, \dots, M$ and extending the definition of quasi-partitions to \mbox{$P_{1:k+1} = \left( \left( W_{1:k}^0, \, W_{k+1}^0 \right), \dots, \left( W_{1:k}^M, \, W_{k+1}^M \right)\right)$} we can relabel the sums in (\ref{eq:kk1_diffA}) to yield
\begin{equation}
\frac{\delta^{k+1} F}{\delta Z_{k+1} \delta Z_{k} \cdots \delta Z_{1}} 
 = \left[ \prod_{i=k+2}^s C_i(\dotprod{c_i}{g_i}) \right]  \sum_{ P_{1:k+1} \in \mathcal{P}_{1:k+1}} \left[ \prod_{i=1}^{k+1}\frac{\delta C_i}{\delta W^0_i} \right] \frac{\delta G^1}{\delta W_{1:k+1}^1} \cdots \dfrac{\delta G^M}{\delta W_{1:k+1}^M}. 
\label{eq:diff_gen_rule_k1A}
\end{equation}

With the general form (\ref{eq:diff_gen_rule_k1A}) for the $k+1$ order differential, the differentiation of $F[g_{1:s}, u]$ for $s$ sensors is compactly written as
\begin{equation}
\frac{\delta^s F}{\delta Z_s \delta Z_{s-1} \cdots \delta Z_{1}} 
 = \sum_{ P_{1:s} \in \mathcal{P}_{1:s}} \left[ \prod_{i=1}^s\frac{\delta C_i}{\delta W^0_i} \right] \frac{\delta G^1}{\delta W_{1:s}^1} \cdots \dfrac{\delta G^M}{\delta W_{1:s}^M}. 
\label{eq:diff_gen_rule_sA}
\end{equation}
Next, we focus on the clutter and Bernoulli PGFl differentials. Recall that $\Gamma_i = \prod_{\vect{z} \in Z_i}  c_i(\vect{z})$ and $C_i(\cdot)^{(n)}$ is the $n$-th differential of the pgf of the cardinality of the clutter process and $c_i(\cdot)$ denotes the clutter pdf of sensor $i$. Additionally, we employ the convention that whenever $\card{W^0_i}=0$, $\prod_{\vect{z}\in W^0_i} (\cdot) =1$. From \cite{bibl:nannuru_MS_CPHD_2016}, the derivative of the clutter pgf is given by 
\begin{align}
\frac{\delta }{\delta W^0_i} C_i(\left\langle c_i,g_i \right\rangle ) &= C_i^{(\card{W^0_i})}(\left\langle c_i,g_i \right\rangle )\prod_{\vect{z} \in W^0_i}c_i(\vect{z}) \nonumber \\
&= C_i^{(\card{W^0_i})}(\left\langle c_i,g_i \right\rangle )\frac{\Gamma_i}{\prod_{\vect{z} \in Z_i \setminus W^0_i}c_i(\vect{z})}.
\label{eq:clutter_pgf_difA}
\end{align}

For a given measurement $\vect{z}_i$ from the $i$-th sensor, by differentiating $G^j[u \textstyle{\prod_{i=1}^s \phi_{g_i}}]$ in $g_i$ with respect to the set $W_i^j=\lbrace \vect{z}_i \rbrace $ we obtain
\begin{align}
 \frac{\delta}{\delta W_i^j} G^j[u \textstyle{\prod_{l=1}^s \phi_{g_l}}] 
& =\frac{\delta}{\delta W_i^j} \left[ 1-r^{(j)} + r^{(j)} \dotprod{p^{(j)}}{u \textstyle{\prod_{l=1}^s} \phi_{g_l}} \right] \nonumber \\
& =r^{(j)}  \int u(\vect{x})p^{(j)}(\vect{x}) p_{i,D}(\vect{x}) h_i(\vect{z}_i^{} \vert \vect{x})\prod_{\substack{l=1 \\ l\neq i}}^s \phi_{g_l}(\vect{x}) d\vect{x}, \nonumber
\end{align}
\noindent where we employed the differentiation rule for a linear functional \cite[Eq. 11.197]{bibl:mahler_book2007}. Let $T_{W_{1:s}}  = \lbrace (i,l) \vert \vect{z}_i^l \in W_{i} \: \forall \: i=1,\dots, s \rbrace$ denote the set of sensor and measurement indices for all measurements in $W_{1:s}$. Whenever $W^j_{1:s} \neq \emptyset_{1:s}$, the differentiation of the Bernoulli PGFl leads to
\begin{equation}
\frac{\delta }{\delta W^j_{1:s}} G^j[u \textstyle{\prod_{i=1}^s} \phi_{g_i}]  =
\displaystyle{ r^{(j)} \int u(\vect{x}) p^{(j)}(\vect{x})} 
 \; \prod_{ \mathclap{(i,l) \in T_{W_{1:s}^j}}} \; p_{i,D}(\vect{x}) h_i(\vect{z}_i^{l} \vert \vect{x})\: \prod_{ \mathclap{(i,*) \notin T_{W_{1:s}^j}}} \: \phi_{g_i}(\vect{x}) d\vect{x}.  
\label{eq:giffG_compA}
\end{equation} 
\noindent  In addition, note the following equality: $\prod_{i=1}^s \prod_{\vect{z}\in Z_i \setminus W_i^0} c_i(\vect{z}) = \prod_{j=1}^M \prod_{(i,l)\in T_{W_{1:s}^j}} c_i(\vect{z}_i^l)$.    

Introducing $\varphi_{W_{1:s}^j}^j[ g_{1:s}, u]$ as 
\begin{equation}
\varphi_{W_{1:s}^j}^j[ g_{1:s}, u] \triangleq  \frac{\frac{\delta }{\delta W^j_{1:s}} G^j[u \textstyle{\prod_{i=1}^s} \phi_{g_i}] }{ \prod_{(i,l) \in T_{W^j_{1:s}}} c_i^{}(\vect{z}_i^l) }  
\label{eq:varphi_def_gA}
\end{equation}
and with the result of (\ref{eq:clutter_pgf_difA}) we can write (\ref{eq:diff_gen_rule_sA}) as
\begin{equation}
\frac{\delta^s F}{\delta Z_s \delta Z_{s-1} \cdots \delta Z_{1}}[g_{1:s},u] =  \left[ \prod_{i=1}^s\Gamma_i \right] 
 \sum_{ P_{1:s} \in \mathcal{P}_{1:s}} \left[ \prod_{i=1}^s C_i^{(\card{W^0_i})} (\dotprod{c_i}{g_i}) \right] \left[ \prod_{j=1}^M \varphi_{W_{1:s}^j}^j[ g_{1:s}, u] \right]. 
\label{eq:diff_gen_rule_s2A}
\end{equation}
Additionally, let $ \mathcal{K}_{P_{1:s}} \triangleq \prod_{i=1}^s C_i^{(\card{W^0_i})} (0)$ and $\varphi_{W_{1:s}^j}^j[u] \triangleq \varphi_{W_{1:s}^j}^j[ 0,\dots, 0, u]$. Then evaluating (\ref{eq:diff_gen_rule_s2A}) in $g_1 = 0,\cdots, g_s =0 $ yields 
\begin{equation}
\frac{\delta F}{\delta Z_{1:s}}[0,\dots,0,u]=
 \left[ \prod_{i=1}^s\Gamma_i \right]  \sum_{ P_{1:s} \in \mathcal{P}_{1:s}} \mathcal{K}_{P_{1:s}} \left[ \prod_{j=1}^M \varphi_{W_{1:s}^j}^j[u] \right], \nonumber
 \label{eq:func_deriv_g_def2A}
\end{equation}
\noindent which represents the main result (\ref{eq:func_deriv_g_def}) of Lemma \ref{the:derivated_funcl}.

\qed

\section{Proof of Theorem \ref{the:ms_mber_phd}} \label{app:proof_updated_phd}
The PHD function corresponding to the updated posterior, as given by (\ref{eq:phd_def}) and (\ref{eq:up_pgfl_def}), is   
\begin{equation}
 D_{k+1 \vert k+1}(\vect{x}) = \left. \frac{\frac{\delta^{s+1}  F}{\delta \vect{x} \delta Z_{1,k+1} \cdots \delta Z_{s,k+1}}[0,0,\dots,0,u]}{\frac{\delta^s F}{\delta Z_{1,k+1} \cdots \delta Z_{s,k+1}}[0,0,\dots,0,1]}\right\vert_{u=1}. \nonumber 
 \label{eq:up_phd_proof}
\end{equation}
\noindent The derivative of the functional $F[\cdot]$ with respect to the test function $u(\cdot)$ yields
\begin{align}
 \left.\frac{\delta^2 F}{\delta \vect{x} \delta Z_{1:s,k+1}}[0,\dots,0,u] \right\vert_{u=1}  
& =\left.\displaystyle \left[ \prod_{i=1}^s\Gamma_i \right] \sum_{ {P_{1:s} \in \mathcal{P}_{1:s}}} \mathcal{K}_{P_{1:s}} \frac{\delta}{\delta \vect{x}} \left\lbrace \prod_{j=1}^M \varphi_{W_{1:s}^j}^j[u] \right\rbrace \right\vert_{u=1} \nonumber \\
& =\left[ \prod_{i=1}^s\Gamma_i \right] \; \; \: \sum_{\mathclap{P_{1:s} \in \mathcal{P}_{1:s}}} \; \mathcal{K}_{P_{1:s}} \hspace{-1.5mm}\left[ \prod_{j=1}^M  \varphi^j_{W_{1:s}^j}[1]\right] \sum_{j = 1}^M \rho^{j}_{W_{1:s}^j}(\vect{x}) p^{(j)}(\vect{x}),
\label{eq:func_deriv_g_u}
\end{align}

\noindent where $\rho^{j}_{W_{1:s}^j}(\vect{x})$ is defined in (\ref{eq:rho_def_main}) and $\varphi^j_{W_{1:s}}[1]$ is assumed non-zero for $\forall$ $j$ and $\forall$ $W_{1:s}$.
\qed

\section{Greedy subset selection algorithm} \label{app:subset_sel}
In Algorithm \ref{alg:subset_sel}, we present the pseudo-code for the greedy selection algorithm employed to select at most $W_{\text{max}}+1$ best-scoring subsets for each of the $M_{k+1 \vert k}$ predicted Bernoulli components. The inputs of the algorithm are given by the parameters of the predicted set of Bernoulli components, the sensor measurements and the maximum number of subsets $W_{\text{max}}$. The algorithm outputs the multi-sensor subsets $ W_{1:s}^{j,l}$ with scores $\beta_{1:s}^{j,l}$ for $l=1,\dots, n_s^j$ and each $j=1,\dots, M_{k+1\vert k}$. Note the independent processing of the predicted Bernoulli components. For each Bernoulli component, the sensors are processed sequentially (line $4$). The $m_i$ measurements of the $i$-th sensor are used to branch the existing $L$ partial subsets into $L \times (m_i+1)$ candidate subsets $U$ (lines $6-15$) and evaluated via $\beta_{1:i}^j$ (line $13$). The path corresponding to the all empty subset $U(1)$ is always retained (line $17$). The non-empty subsets are sorted (line $18$) in decreasing order of their scores $w$, while the sorting function $\texttt{sort}$ returns the sorting indices. Finally, at most $W_{\text{max}}$ subsets are retained from the non-empty candidate subsets $U(2), \dots, U((m_i+1)L)$ on lines $19-20$. The complexity of Algorithm \ref{alg:subset_sel} is $\mathcal{O}(M_{k+1\vert k}W_{\text{max}}\,\sum_{i=1}^s m_i)$, where the complexity of the sorting operation was considered negligible with respect to the complexity of $L \times (m_i+1)$ scoring operations, i.e, computation of $\varphi_{U}^{(j)}[1]$ which depends on the implementation type (Kalman, EKF, UKF or particle filter).      
\begin{figure}[t]
{\centering
\begin{minipage}{1.\linewidth}
\begin{algorithm}[H]
\caption{Greedy subset selection}\label{alg:subset_sel}
\begin{algorithmic}[1]
\Function {\texttt{Greedy\_subset\_selection}}{}
\Statex  $( \lbrace r_{k+1 \vert k}^{(j)}, p_{k+1 \vert k}^{(j)}(\vect{x}) \rbrace_{j=1}^{M_{k+1 \vert k}}, \lbrace Z_{i,k+1} \rbrace_{i=1}^s$,$W_{\text{max}} )$
\For{$j\leftarrow  1$ to $M_{k+1 \vert k}$}
\State Initialize path: $L\leftarrow1$, $W_0^{j,1} \leftarrow [\,]$
\For{$i \leftarrow  1$ to $s$}
\State $U \leftarrow [\,]$, $w \leftarrow [\,]$


\For{$n \leftarrow 0$ to $m_i$}

\For{$l \leftarrow 1$ to $L$}


\If{$n=0$}
\State $U(l+nL) \leftarrow \left(W_{1:i-1}^{j,l} ,\, \emptyset \right)$
\Else 

\State $U(l+nL) \leftarrow \left(W_{1:i-1}^{j,l} ,\, \{\vect{z}_{i}^n \}\right)$

\EndIf
\State $w(l+nL) \leftarrow \varphi_{U(l+nL)}^{(j)}[1] $
\EndFor
\EndFor
\State $L \leftarrow \texttt{min}(W_{\text{max}}, (m_i+1)L-1)$
\State $\beta_{1:i}^{j,1} \leftarrow w(1)$, $W_{1:i}^{j,1} \leftarrow U(1)$
\State $\text{sort\_idx} \leftarrow \texttt{sort}(w(2), \dots, w(end) )$
\State $W_{1:i}^{j,l+1} \leftarrow U(\text{sort\_idx}(l)+1)$ for $l=1, \dots, L$
\State $\beta_{1:i}^{j,l+1} \leftarrow w(\text{sort\_idx}(l)+1)$ for $l=1, \dots, L$
\EndFor
\State $n_s^j \leftarrow L+1$
\EndFor
  \State \Return $ \lbrace ( \beta_{1:s}^{j,l}, W_{1:s}^{j,l} ) \vert l=1, \dots, n_s^j \rbrace$ for 
  \Statex  \hspace{4.2cm} $j=1, \dots, M_{k+1 \vert k}$
\EndFunction
\end{algorithmic}
\end{algorithm} 
\end{minipage}
\par
}
\end{figure}

\section{Greedy partition selection algorithm} \label{app:part_sel}
        
\begin{figure}[!t]
{\centering
\begin{minipage}{1.\linewidth}
\begin{algorithm}[H]
\caption{Greedy partition selection algorithm}\label{alg:part_sel}
\begin{algorithmic}[1]
\Function{ \texttt{Greedy\_partition\_selection }}{}
\Statex $ ( \lbrace ( \beta_{1:s}^{j,l}, W_{1:s}^{j,l} ) \vert l=1, \dots, n_s^j \rbrace_{j=1}^{M_{k+1\vert k}}$,$P_{\text{max}} )$
\State Initialize partitions: $n_P \leftarrow 1$, $P_0^1\leftarrow [\,]$, $\alpha_0^1 \leftarrow 1$
\For{$j\leftarrow  1$ to $M_{k+1 \vert k}$}
\State Initialize path: $w\leftarrow [\,]$, $Q \leftarrow [\,]$, $n \leftarrow 1$
\For{$i \leftarrow  1$ to $n_P$}
\For{$l \leftarrow 1$ to $n_s^j$}
\If{$\neg \texttt{overlap}\left(P_{1:j-1}^i, W_{1:s}^{j,l}\right)$}
\State $Q(n) \leftarrow \left(P_{1:j-1}^i, W_{1:s}^{j,l}\right) $
\State $w(n) \leftarrow \alpha_{1:j-1}^i \beta_{1:s}^{j,l} $
\State $n \leftarrow n+1$
\EndIf
\EndFor
\EndFor
\State $n_P \leftarrow \texttt{min}(P_{\text{max}},n-1)$
\State $\text{sort\_idx} \leftarrow \texttt{sort}(w(1), \dots, w(n-1) )$
\State $P_{1:j}^i \leftarrow Q(\text{sort\_idx}(i))$ for $i=1, \dots, n_P$
\State $\alpha_{1:j}^i \leftarrow w(\text{sort\_idx}(i))$ for $i=1, \dots, n_P$
\EndFor
  \State \Return $ \lbrace ( \alpha_{1:M_{k+1 \vert k}}^i, P_{1:M_{k+1 \vert k}}^i ) \rbrace_{i=1}^{n_P}$
\EndFunction
\end{algorithmic}
\end{algorithm} 
\end{minipage}
\par
}
\end{figure}
\begin{figure}[!t]
{\centering
\begin{minipage}{1.\linewidth}
\begin{algorithm}[H]
\caption{Pseudo-code of function evaluating the possible overlap between a partition and a subset}\label{alg:overlap}
\begin{algorithmic}[1]
\Function{ \texttt{overlap}}{ $P_{1:j-1} = \left( Q_{1:s}^{(1)}, \dots,Q_{1:s}^{(j-1)} \right)$, $W_{1:s}^{j} = \left( W_1, \dots, W_s\right)$}
\State $\text{flag} \leftarrow 0$, $n=0$
\While{ $(\neg \text{flag}) \wedge (n<j-1)$ }
\State $n \leftarrow n+1$
\For{$i \leftarrow 1$ to $s$}
\If{ $ (W_i \neq \emptyset) \wedge (W_i = Q_i^{(n)}) $}
\State $\text{flag} \leftarrow 1$
\EndIf 
\EndFor
\EndWhile
\State \Return $\text{flag}$
\EndFunction
\end{algorithmic}
\end{algorithm} 
\end{minipage}
\par
}
\end{figure}

In Algorithm \ref{alg:part_sel}, we present the pseudo-code for the greedy selection algorithm employed to select at most $P_{\text{max}}$ best-scoring partitions. The inputs of the algorithm are the multi-sensor subsets and their associated scores for each predicted Bernoulli component and the maximum number of desired partitions $P_{\text{max}}$. The algorithm returns the set of selected partitions with their scores $\alpha_{1:M_{k+1\vert k}}$. The predicted Bernoulli components are sequentially processed (line $3$) while existing selected paths $P_{1:j-1}$ are branched into candidate paths $P_{1:j}$ by appending multi-sensor subsets from the $j$-th Bernoulli component (lines $5-13$). Note that only a non-overlapping multi-sensor subset is added to given partition, i.e., that does not have any measurements in common with the subsets already contained in the respective partition (line $7$). In Algorithm \ref{alg:overlap}, we present the pseudo-code for the algorithm that verifies the overlap condition. The candidate partitions are scored at line $9$ and a sorting operation is employed to retain at most $P_{\text{max}}$ high-scoring partitions (lines $16-17$). In Algorithm \ref{alg:overlap} on line $6$, we verify if the $i$-th sensor measurement $\vect{z}_i$ contained in $W_{1:s}$ is also contained in the multi-sensor subset $Q_{1:s}^{(n)}$. The worst case computational complexity of Algorithm \ref{alg:overlap} is $\mathcal{O}(sM_{k+1 \vert k})$, which leads to a complexity of $\mathcal{O}(P_{\text{max}}\,W_{\text{max}}\,s\, M_{k+1 \vert k}^2)$ for Algorithm \ref{alg:part_sel}. Note that the computational complexity of the sorting operation is negligible with respect to the complexity of $P_{\text{max}}W_{\text{max}}$ repeated calls to the function $\texttt{overlap}$.

\section{Truncated MS-CPHD filter}
\label{app:tcphd}
In this appendix, we present the Truncated MS-CPHD (MS-TCPHD) filter. The MS-TCPHD filter is obtained by modifying the update step of the MS-CPHD filter while keeping the same prediction step. As described in \cite{bibl:nannuru_MS_CPHD_2016}, the MS-CPHD filter employs the following update equation
\begin{equation}
D_{k+1 \vert k+1}^{\text{CPHD}}(\vect{x}) = \left( \alpha_0 \prod_{i=1}^s (1-p_{i,D}(\vect{x})) + \sum_{P \in \mathcal{P}} \alpha_P \sum_{W_{} \in P}\rho_{W_{}}(\vect{x}) \right)D^{\text{CPHD}}_{k+1 \vert k}(\vect{x}), 
\label{eq:ms_CPHD_nannuru}
\end{equation} 
\noindent where $D^{\text{CPHD}}_{k+1 \vert k}(\vect{x})$ is the normalized (i.e., $\int D^{\text{CPHD}}_{k+1 \vert k}(\vect{x}) d\vect{x}=1$) predicted PHD function and the expressions of $\alpha_0$, $\alpha_P$ and $\rho_W(\cdot)$ are found in \cite[Eqs. (20-23)]{bibl:nannuru_MS_CPHD_2016}. Considering the predicted PHD function as $D_{k+1 \vert k}^{\text{CPHD}}(\vect{x}) = \sum_{j=1}^{J_{k+1\vert k}} \omega_{k+1\vert k}^{(j)}  \mathcal{N}(\vect{x};\mu_{k+1\vert k}^{(j)}, \matg{\Sigma}_{k+1\vert k}^{(j)})$, the MS-CPHD filter of \cite{bibl:nannuru_MS_CPHD_2016} updates each Gaussian component with all of the subsets contained in $P$ via $\rho_{W_{}}(\cdot)$.

For the MS-TCPHD filter, we employ similar greedy subset and partition formation mechanisms as in~\cite[Sec.V]{bibl:nannuru_MS_CPHD_2016}, where the Gaussian mixture components represent potential targets. Consequently, the score $\beta^j(W_{})$ for updating the $j$-th Gaussian mixture component with the non-empty subset $W_{}$ is 
\begin{equation}
\beta^j(W_{}) \triangleq \frac{\int \omega_{k+1\vert k}^{(j)} \mathcal{N}(\vect{x};\mu_{k+1\vert k}^{(j)}, \matg{\Sigma}_{k+1\vert k}^{(j)}) \left( \prod_{(i,l) \in T_{W_{}} } p_{i,D} (\vect{x}) h_i(\vect{z}_i^l \vert \vect{x}) \right) \prod_{i:(i,*)\notin T_{W_{}}} (1-p_{i,D} (\vect{x})) d \vect{x} }{\prod_{(i,l)\in T_{W_{}}} c_i(\vect{z}_i^l)}. 
\label{eq:beta_TCPHD}
\end{equation}
The greedy subset selection procedure yields a set of high-scoring subsets associated with each Gaussian component $j=1, \dots, J_{k+1 \vert k}$. In the following, a subset selected for the $j$-th Gaussian component will be denoted with $W_{}^j$ while $V$ is the clutter subset. The greedy partition selection procedure generates partitions $P =\{V, W_{}^1, \dots, W_{}^{J_{k+1 \vert k}} \} $ with high scores $\prod_{W^j_{} \in P} \beta^j(W_{}^j)$. For a given partition $P $, the MS-TCPHD filter updates the $j$-th mixture component with its associated non-empty subset $W^j_{}$. In contrast, the MS-CPHD filter updates the $j$-th mixture component with all $W^i_{}$ for $i=1, \dots, J_{k+1 \vert k}$. For well separated targets, the truncated update scheme of the MS-TCPHD filter is justified since the measurement subsets selected by a mixture component do not significantly correlate with the other mixture components. Hence, for a non-empty subset $W_{}^j$ the subset score $d_{W_{}^j} = \sum_{i=1}^{J_{k+1 \vert k}}\beta^i(W_{}^j)$ of \cite[Eq. (19)]{bibl:nannuru_MS_CPHD_2016} is approximated as $\hat{d}_{W_{}^j}= \beta^j(W_{}^j)$ in the MS-TCPHD filter. The subset scores $\hat{d}_{W_{}^j}$ are then used to obtain the approximate values $\hat{\alpha}_0$ and $\hat{\alpha}_P$ via equations (21) and (22) of \cite{bibl:nannuru_MS_CPHD_2016}.

In the MS-TCPHD filter, the updated PHD function is given by
\begin{equation}
D_{k+1 \vert k+1}^{\text{TCPHD}}(\vect{x})  = \hat{\alpha}_0 D_{k+1 \vert k}^{\text{TCPHD}}(\vect{x}) \prod_{i=1}^s (1-p_{i,D}(\vect{x}))
 +  \sum_{{\substack{P \in \mathcal{P}}}}  \hat{\alpha}_P \left( \sum_{W^j_{} \in P} \omega_{k+1\vert k}^{(j)}\; \hat{\rho}^j_{W_{}^j}(\vect{x}) \mathcal{N}(\vect{x};\mu_{k+1\vert k}^{(j)}, \matg{\Sigma}_{k+1\vert k}^{(j)}) \right), 
\label{eq:trunc_cphd}
\end{equation} 
where 
\begin{equation}
\hat{\rho}^j_{W^j}(\vect{x}) = \frac{ \left( \prod_{(i,l) \in T_{W^j} } p_{i,D} (\vect{x}) h_i(\vect{z}_i^l \vert \vect{x}) \right) \prod_{i:(i,*)\notin T_{W^j}} (1-p_{i,D} (\vect{x}))  }{\int \omega_{k+1\vert k}^{(j)} \mathcal{N}(\vect{x};\mu_{k+1\vert k}^{(j)}, \matg{\Sigma}_{k+1\vert k}^{(j)}) \left( \prod_{(i,l) \in T_{W^j} } p_{i,D} (\vect{x}) h_i(\vect{z}_i^l \vert \vect{x}) \right) \prod_{i:(i,*)\notin T_{W^j}} (1-p_{i,D} (\vect{x})) d \vect{x}}.
\end{equation}
Observe from (\ref{eq:trunc_cphd}) that under a partition $P$ and for a non-empty subset $W^j \in P$, the $j$-th PHD mixture component is updated with its associated measurement subset via $\hat{\rho}^j_{W_{}^j}(\vect{x})$. Given the approximate values $\hat{d}_{W^j}$, the cardinality distribution of the MS-TCPHD filter is updated via equation (24) of \cite{bibl:nannuru_MS_CPHD_2016}.
\ifCLASSOPTIONcaptionsoff
  \newpage
\fi



%

\bibliographystyle{IEEEbib}
\bibliography{biblio_alex_sp2}

%

%
%
%




\end{document}